# Electrons and polarons at oxide interfaces explored by soft-X-ray ARPES

Vladimir N. Strocov[1], Claudia Cancellieri[2] and Andrey S. Mishchenko[3]

Soft-X-ray ARPES (SX-ARPES) with its enhanced probing depth and chemical specificity allows access to fundamental electronic structure characteristics – momentum-resolved spectral function, band structure, Fermi surface – of systems difficult and even impossible for the conventional ARPES such as three-dimensional materials, buried interfaces and impurities. After a recap of the spectroscopic abilities of SX-ARPES, we review its applications to oxide interfaces, focusing on the paradigm $LaAlO_3/SrTiO_3$ interface. Resonant SX-ARPES at the Ti *L*-edge accentuates photoemission response of the mobile interface electrons and exposes their $d_{xy}$-, $d_{yz}$- and $d_{xz}$-derived subbands forming the Fermi surface in the interface quantum well. After a recap of the electron-phonon interaction physics, we demonstrate that peak-dip-hump structure of the experimental spectral function manifests the Holstein-type large polaron nature of the interface charge carriers, explaining their fundamentally reduced mobility. Coupling of the charge carriers to polar soft phonon modes defines dramatic drop of mobility with temperature. Oxygen deficiency adds another dimension to the rich physics of $LaAlO_3/SrTiO_3$ resulting from co-existence of mobile and localized electrons introduced by oxygen vacancies. Oxygen deficiency allows tuning of the polaronic coupling and thus mobility of the charge carriers, as well as of interfacial ferromagnetism connected with various atomic configurations of the vacancies. Finally, we discuss spectroscopic evidence of phase separation at the $LaAlO_3/SrTiO_3$ interface. Concluding, we put prospects of SX-ARPES for complex heterostructures, spin-resolving experiments opening the totally unexplored field of interfacial spin structure, and in-operando field-effect experiments paving the way towards device applications of the reach physics of oxide interfaces.

[1] Swiss Light Source, Paul Scherrer Institute, CH-5232 Villigen-PSI, Switzerland
[2] EMPA, Swiss Federal Laboratories for Materials Science & Technology, Ueberlandstrasse 129, CH-8600 Duebendorf, Switzerland
[3] RIKEN Center for Emergent Matter Science (CEMS), 2-1 Hirosawa, Wako, Saitama 351-0198, Japan

# Contents

# 1 Soft-X-ray ARPES: From bulk materials to interfaces and impurities

Angle-resolved photoelectron spectroscopy (ARPES) as the unique experimental technique delivering direct information about electronic structure of crystalline solids resolved in electron energy $E$ and momentum **k**. While the fundamentals of ARPES as well as characteristic spectroscopic properties of this technique in different photon energy ranges have been discussed in Introduction to this volume, in this Chapter we will focus on spectroscopic abilities of soft-X-ray ARPES (SX-ARPES). Combining the **k** resolution with enhanced probing depth and chemical specificity, this technique is ideally suited for buried interface and impurity systems which are in the core of nowadays and future electronic and spintronic devices.

## 1.1 Virtues and challenges of soft-X-ray ARPES

SX-ARPES as a spectroscopic technique exploiting the region of photon energies $hv$ around 1 keV has been pioneered at SPRing-8 (Sekiyama 2004; Suga 2004; Suga and Tusche 2015) and has recently been boosted with advent of synchrotron sources delivering high photon flux in this energy range such as the Swiss Light Source (SLS) and Diamond Light Source (DLS). SX-ARPES features a few fundamental advantages compared to the conventional VUV-ARPES with its $hv$ region around 20-100 eV:

*Probing depth.* The "universal curve" of the photoelectron attenuation length $\lambda$ [see, for example, Powell et al. (1999)] shows that VUV-ARPES is characterised by a probing depth of a few Å, which makes this technique extremely surface sensitive, and limits its applications basically to atomically clean surfaces. The increase of $\lambda$ in the SX-ARPES energy range enhances bulk sensitivity of the ARPES experiment as well as enables access to systems buried behind a few surface layers, often without any surface preparation. SX-ARPES is therefore highly relevant for real-world materials of importance in modern solid-state technology.

*3D momentum resolution.* The photoemission (PE) final state confinement within $\lambda$ results, by the Heisenberg uncertainty principle, in intrinsic broadening of the corresponding surface-perpendicular momentum $k_\perp$ defined by $\Delta k_\perp = \lambda^{-1}$. The increase of $\lambda$ at higher energies results therefore in improvement of the intrinsic $k_\perp$ resolution (Strocov 2003; Strocov et al. 2012). In combination with free-electron final-state dispersion, this sharpens the definition of $k_\perp$ and thus of the full 3D momentum to enable precise determination of the valence band (VB) dispersions $E(\mathbf{k})$ in 3D materials [see examples below as well as Eguchi et al. (2009); Strocov et al. (2012); Lev et al. (2015); etc] problematic for the VUV-ARPES.

*Resonant photoemission spectroscopy (ResPES).* The SX-ARPES energy range goes through a number of important absorption edges including the *L*-edges of transition metals and *M*-edges of rear-earths. This allows resonant excitation of their valence *d*- and *f*-states which play the crucial role in the physics of strong electron correlations. In this way resonant SX-ARPES not only reveals the elemental composition of the VB (Olson et al. 1996; Molodtsov et al. 1997) but also can be used to highlight the signal from buried systems.

A few applications of SX-ARPES exploiting the above virtues of enhanced probing depth, 3D momentum definition and chemical specificity will be illustrated later.

Realization of these SX-ARPES virtues comes however with a few severe challenges:

*Cross-section problem.* The main challenge of SX-ARPES, until now severely impeding its practical use, is that the VB photoexcitation cross-section reduces compared to the VUV energy range typically by 2-3 orders of magnitude (Yeh and Lindau 1985) because overlap of the rapidly oscillating high-energy final states with smooth valence states reduces essentially to the small ion core region (Solterbeck et al. 1997). Such a dramatic signal loss has to be compensated by high flux of incoming photons, which requires the most advanced synchrotron radiation sources and beamline instrumentation, as well as efficient photoelectron detectors.

*Electron-phonon interaction.* In the soft-X-ray energy range the photoelectron wavelengths become comparable with phononic displacements of atoms in the unit cell, which relaxes the **k**-conserving dipole selection rules. This electron-phonon interaction effect manifests itself as momentum broadening of the spectral structures as well as transfer of the coherent spectral weight $I^{coh}$ into an incoherent background reflecting the matrix-element weighted **k**-integrated density of states (DOS) (Venturini et al. 2008; Braun et al. 2013). The reduction of $I^{coh}$ relative to that at zero temperature $I^{coh}_{T=0}$ can be expressed, in the first approximation, as $I^{coh} = W(T)I^{coh}_{T=0}$, where $W(T)$ is the photoemission Debye-Waller factor $W(T) = e^{-\Delta G^2 U_0^2(T)}$ with $\Delta G \propto \sqrt{h\nu}$ expressing the momentum transfer between the initial and final state and $U_0$ the *rms* thermal atomic displacement (Papp et al. 2011). Cryogenic sample cooling, ultimately towards liquid-He temperatures, is therefore imperative to achieve adequate **k**-resolution of SX-ARPES. More on the temperature effects can be found in Chapter 7.

Further increase of $h\nu$ into the multi-keV energy range pushes $\lambda$ to more than 50 Å (Powell et al. 1999). The corresponding hard-X-ray ARPES (HX-ARPES, see Chapter 7) has however to stand progressive reduction of the VB cross-section and loss of the coherent spectral weight with energy. In addition, recoil from high-energy photoelectrons – essentially, emission of phonons back in the lattice (Hofmann et al. 2002) – smears photoelectron **k** and energy (Suga and Sekiyama 2009). Therefore, the **k**-resolving abilities of HX-ARPES seem to stay with high-Z materials like W or materials with stiff covalent bonds like GaAs (Papp et al. 2011; Gray et al. 2011; Fadley 2012) where the temperature and recoil effects are small. SX-ARPES appears therefore as a winning combination of enhanced probing depth with sufficient VB cross-section and **k**-resolution.

## 1.2 Experimental technique

The cross-section problem of SX-ARPES requires synchrotron radiation beamlines delivering high photon flux. At the time of writing, the worldwide highest brilliance soft-X-ray source was the ADRESS beamline at the SLS (for details see Strocov et al. 2010). The beamline delivers soft-X-ray radiation with variable linear and circular polarizations in $h\nu$ range from 300 to 1600 eV with an ultimate resolving power $E/\Delta E$ of 33'000 at 1 keV. By virtue of the SLS ring energy 2.4 GeV optimal for the soft-X-ray range and the beamline design optimized for high transmission, the photon flux tops up $10^{13}$ photons/s/0.01%BW. Apart from SX-ARPES, the ADRESS beamline hosts a Resonant X-ray Scattering (RIXS) facility delivering complementary information on charge-neutral excitations (see Chapter 11). Dedicated SX-ARPES facilities are also available at SPRing-8, DLS, PETRA-III, and more are coming soon at MAX-IV and NSLS-II.

Another factor to increase the PE signal is a grazing-incidence experimental geometry dictated by interplay of relatively large X-ray penetration depth $d$ with relatively small photoelectron escape depth $\lambda$ [Henke 1972; Fadley et al. 2003; Strocov 2013). Fig. 1 (*a*) shows the normal-emission photocurrent $I_{PE}$ as a function of X-ray grazing incidence angle $\alpha$ calculated near the ends of the soft-X-ray energy interval for the valence states of two paradigm materials Cu and GaAs (Strocov 2013). $I_{PE}(\alpha)$ sharply increases when going to grazing $\alpha$, and reaches its maximum near the critical reflection angle $\alpha_c$ when the electromagnetic field concentrates in the near-surface region where the photoelectrons can escape from. Importantly, with the increase of $h\nu$ the $I_{PE}(\alpha)$ maximum shifts to very grazing $\alpha$ around a few degrees. Optimal for the SX-ARPES experiment will be therefore a grazing-incidence geometry (Strocov et al. 2014-1) sketched in Fig. 1 (*b*). Importantly, the sample is rotated to the grazing α around the horizontal axis, because in this case the X-ray footprint on the sample blows up in the vertical plane where the synchrotron beam has much smaller size than in the horizontal. The analyzer slit is oriented in the measurement plane that enables symmetry analysis of the valence states by switching the X-ray polarization between horizontal and vertical.

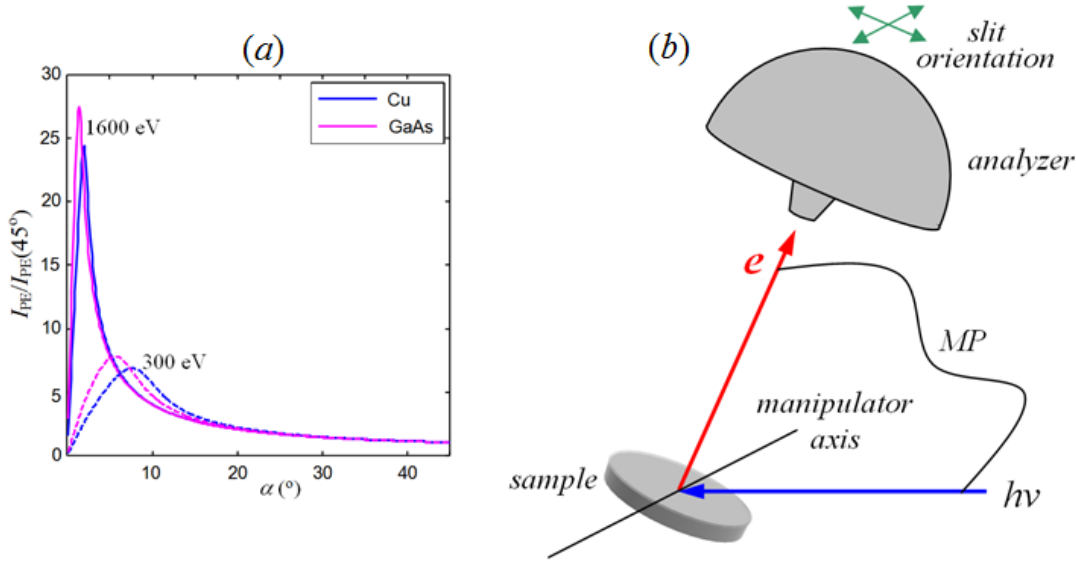

**Fig. 1** (*a*) Dependence of the normal-emission photocurrent $I_{PE}$ on grazing X-ray incidence angle. With increase of *hv* its maximum shifts to more grazing angles (adapted from Strocov 2013); (*b*) Optimized geometry of the SX-ARPES experiment. The grazing X-ray incidence increases the PE signal, horizontal rotation axis reduces the light footprint on the sample, and slit oriented in the measurement plane (MP) allows symmetry analysis of the valence states (Strocov et al. 2014-1).

The SX-ARPES setup at the ADRESS beamline (Strocov et al. 2014-1) operates at $\alpha = 20°$ (at the moment restricted by mechanical constraints) which increases $I_{PE}(\alpha)$ by a factor of ~2 compared to the conventional $\alpha = 45°$. The light footprint on the sample is about 20 x 74 μm. The photoelectron analyzer PHOIBOS-150 from SPECS delivers high angular resolution $\delta\vartheta \sim 0.07°$ FWHM which is particularly important for SX-ARPES because of the corresponding momentum uncertainty $\delta K_{//} = 0.5124\sqrt{E_k}\sin\delta\vartheta$ magnified by high kinetic energy $E_k$. The CARVING manipulator provides three independent angular degrees of freedom, allowing precise navigation in **k**-space. The samples are liquid-He cooled down to 10.7K. The practical combined (beamline plus analyzer) resolving power $E/\Delta E$ of this SX-ARPES facility is up to 20'000. This is the nowadays SX-ARPES state-of-art.

### 1.3 Application examples

We will now flash through a few scientific cases achieved at the SLS which illustrate the above spectroscopic virtues of SX-ARPES. Extending the recent review (Strocov et al. 2014) we will demonstrate that the recent progress in SX-ARPES instrumentation has not only overpowered the dramatic photoexcitation cross-section loss in this energy range but also pushed SX-ARPES to the most photon-hungry cases of buried interfaces and impurities.

#### 1.3.1 Probing deep: Buried interfaces

In the Introduction to this volume we have already seen an illustration of the enhanced probing depth of SX-ARPES with experiments on GaAs(100) capped with a ~10Å thick layer of amorphous As (Kobayashi et al. 2012). A recent example relevant for oxide spintronics is the spin injector heterostructure SiO$_x$/EuO/Si (Lev et al. 2016). EuO is one of the most promising routes for spintronics, for entries see (Miao and Moodera 2015, Caspers et al. 2015). This material, first, delivers almost 100% spin-polarized electrons inviting spin-filter applications and, second, can be integrated into the wide-spread Si technology. Furthermore, unique properties of EuO such as its colossal magnetoresistivity, metal-insulator transition, strong magneto-optic effects, tunability of the magnetic properties by doping or strain, etc. open its ways towards multi-functional devices.

Our experiments on the SiO$_x$/EuO/Si heterostructure sketched in Fig. 2 (*a*) aimed at determination of the EuO/Si band offset, definitive for the spin injector applications. We used *hv* above 1 keV, allowing penetration of photoelectrons excited in the Si substrate through the SiO$_x$/EuO layers. The experimental ARPES intensity image in Fig. 2 (*b*) was measured with *hv* = 1120 eV, tuning $k_\perp$ to the VB maximum of bulk Si in the Γ-point of the Brillouin zone (BZ). We see the dispersionless Eu$^{2+}$ state in the EuO layer buried behind the 17 Å thick layer of SiO$_x$ and, moreover, we recognize the textbook manifold of the light-hole and heavy-hole bands along the ΓKX direction of bulk Si, which is buried behind the SiO$_x$/EuO layer with a total thickness of 30 Å. This is an impressive example of the penetrating ability of SX-ARPES.

The energy difference between the upper edge of the Eu$^{2+}$ multiplet and the VB maximum of Si in the Γ point, Fig. 2 (*c*), directly measures the EuO/Si band offset Δ*E* as 0.8 eV. This value is sufficiently large to make spin injection at the EuO/Si interface immune to noise and, on the other hand, it stays sufficiently small to reduce power consumption. These SX-ARPES results justify thereby the spin injecting functionality of the SiO$_x$/EuO/Si heterostructure.

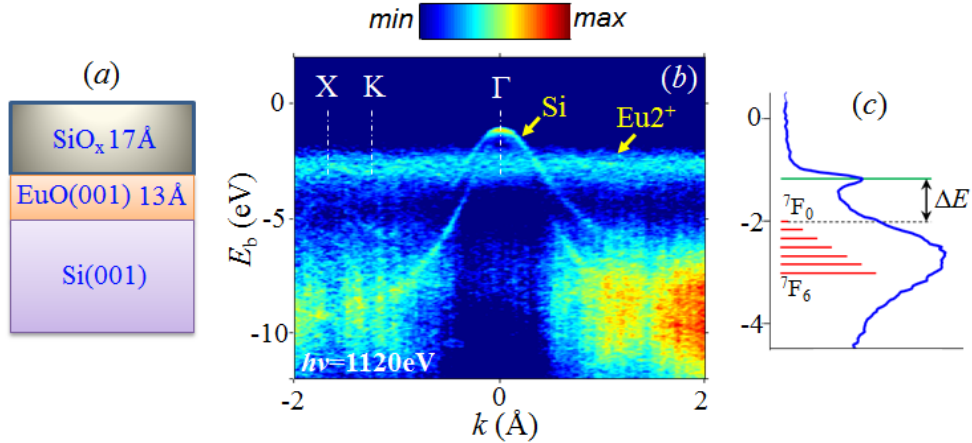

**Fig. 2.** SiOx/EuO/Si spin injector: (*a*) Scheme of the sample; (*b*) SX-ARPES image (angle-integrated component subtracted) measured at *hv* = 1120 eV, which reveals the Eu$^{2+}$ state on top of the three-dimensional *E*(**k**) of bulk Si along the ΓKX direction; (*c*) Identification of the band offset Δ*E* = 0.8 eV which justifies the spin injecting functionality of the SiO$_x$/EuO/Si heterostructure (Lev et al. 2016).

As we will see below, the SX-ARPES enhanced probing depth is essential for experiments on the paradigm TMO interface, buried LAO/STO. In this perspective we should also mention recent studies of dimensionality-tuned electronic structure of LaNiO$_3$/LaAlO$_3$ superlattices (Berner et al. 2015, see Chapter 5) as well as two-dimensional electron gas in a multilayer structure formed by interfacing of the ferromagnetic Mott insulator GdTiO$_3$ with STO (Nemšák et al. 2015).

### 1.3.2 Momentum resolution in 3D: Bulk materials

We will now illustrate the use of the $k_\perp$-momentum resolution of SX-ARPES to explore 3D electronic structure of the typical transition metal dichalcogenide (TMDC) $VSe_2$. The chalcogen-metal-chalcogen layered structure of TMDCs results in their quasi-2D properties (Woolley and Wexler 1977) but the bands derived from the out-of-plane chalcogen orbitals retain pronounced 3D character characterized by $k_\perp$ dispersion ranges of a few eV. Accurate measurements of these dispersions with VUV-ARPES are impeded by large intrinsic $k_\perp$ broadening of the low-energy final states comparable with the perpendicular extension of the BZ (see Strocov et al. 1997; Strocov et al. 2006 and references therein).

Results of SX-ARPES measurements on $VSe_2$ in an $hv$ range around 900 eV are compiled in Fig. 3. The ARPES intensity images (*b*) along two in-plane and out-of-plane directions of the BZ (*a*) clearly show the V 3*d* bands near $E_F$ as well as the Se 4*p* ones deeper in the VB. Statistics of the experimental data is remarkable in view of the photoexcitation cross-section reduction by a factor of ~2000 for the V 3*d* and ~30 for the Se 4*p* states for our excitation energies compared to VUV-ARPES (Yeh and Lindau 1985). Reliable control over $k_\perp$ in the SX-ARPES experiment has allowed slicing the Fermi surface (FS) in different planes, including in- (*d*) and out-of-plane (*c*) ones. The FS topology with its characteristic symmetry pattern is extremely clear in the experimental maps. The textbook clarity of the experimental data compared to the previous VUV-ARPES results (Terashima *et al.* 2003) comes from high $k_\perp$ definition of the high-energy final states. A detailed account of these results, in particular analysis of 3D warping of the FS to form exotic 3D charge density waves, is given in (Strocov et al. 2012).

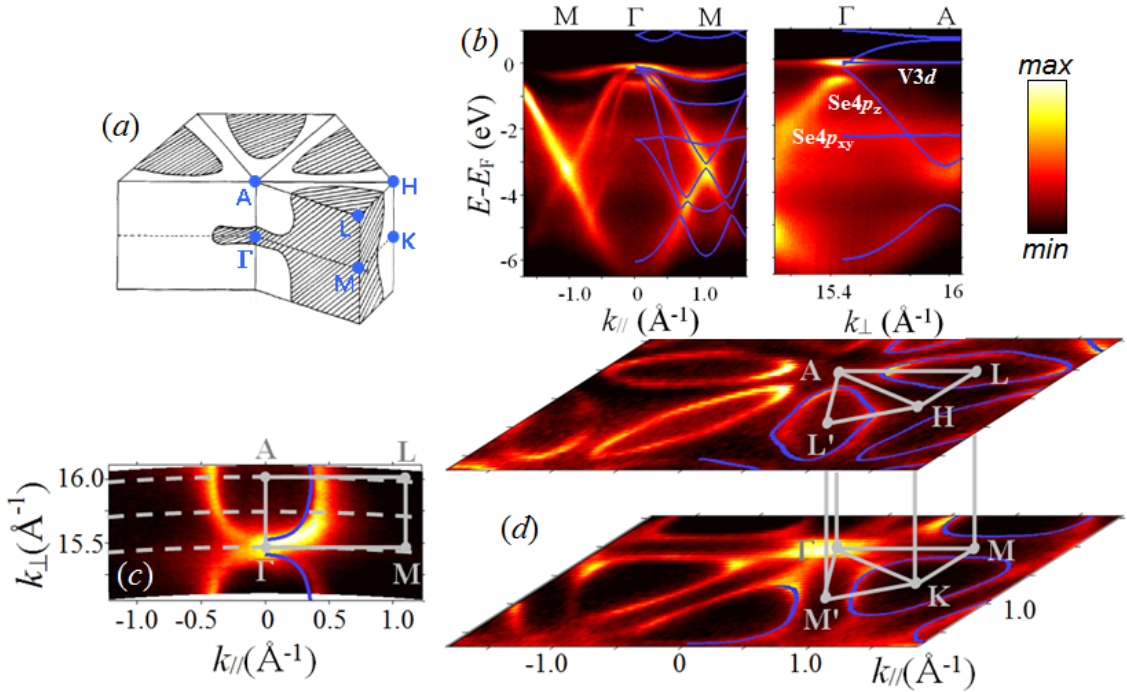

**Fig. 3**. 3D electronic structure of $VSe_2$: (*a*) BZ with its main symmetry lines and inscribed FS; (*b*) ARPES intensity along the M'ΓM line measured at $hv$ = 885 eV, and along the ΓA line under variation of $hv$ to scan $k_\perp$; (*c-d*) experimental FS slices in the (*c*) ΓALM plane (variation of $hv$) and (*d*) ΓKM and ALH planes ($hv$ = 885 and 945 eV, respectively). The clarity of the experimental data manifests $k_\perp$ definition of the high-energy final states (adapted from Strocov et al. 2012).

Of the recent SX-ARPES applications to bulk TMOs, precursors to our main topic of TMO interfaces, we can mention the paradigm manganite $La_{1-x}Sr_xMnO_3$ (LSMO) possessing 3D perovskite structure. Whereas previous VUV-ARPES studies suffered from insufficient $k_\perp$ resolution (J. Krempaský et al. 2008), the SX-ARPES experiments by Lev et al. (2015) employing $h\nu$ around 700 eV have immediately yielded the long-sought-for canonical FS topology consisting of the alternating 3D spheroid electron pockets and cuboid hole pockets. Shadow contours of the experimental FS, reminiscent of the high-$T_c$ cuprates (Damascelli et al. 2003), manifested the rhombohedral lattice distortion of LSMO. This distortion is neutral to the Jahn-Teller effect and thus polaronic coupling, but reduces the DE electron hopping along the Mn-O-Mn bonds. In interplay of the polaronic self-localization with the DE electron delocalization (Millis et al. 1996; Millis 1998; Hartinger et al. 2006), this effect reduces the colossal magnetoresistance $T_c$. Most recently, SX-ARPES has been applied to Ce-doped $CaMnO_3$ (CMO) to reveal shadow contours of the experimental FS as signatures of the orthorhombic lattice distortion, and strong band dependent polaronic coupling (Husanu et al., unpublished).

Of other applications of SX-ARPES to bulk TMOs, we should mention a pioneering study of band structure and FS of $CrO_2$ probed through a layer of amorphous $Cr_2O_3$ inherently building up on its surface (Bisti et al. 2016). Further examples of the 3D electronic structure determination with SX-ARPES in various materials range from layered materials such as $VSe_2$ where 3D nesting of the FS results in 3D charge density waves (Strocov *et al.* 2012) to the conventional high-$T_c$ superconductor $MgB_2$ with dichotomy of the Fermi states in their dimensionality (Sassa et al. 2015), AlNiCo quasicrystals with anisotropic structure stabilization mechanism (Rogalev et al. 2015), strongly correlated ruthenates and pnictides (Razzoli et al. 2012; Razzoli et al. 2015), bandwidth controlled Mott materials (Xu et al. 2014), heavy fermion systems (Höppner et al. 2013), etc. In the study on the 3D pnictide $(Ba_{1-x}K_x)Fe_2As_2$ by Derondeau et al (2017) it was noted that accuracy of the electron effective mass determination is much affected by the $k_\perp$ broadening effects and therefore improves with increase of $h\nu$. Furthermore, we should particularly mention Weil semimetals where high $k_\perp$ definition of SX-ARPES was crucial to identify the 3D cones topologically connecting the surface state arcs (see Lv et al. 2015; Xu et al. 2015-1; Xu et al. 2015-2; Xu et al. 2016; Di Sante et al. 2017; Lv et al. 2017; etc.). In connection with spin physics of 3D materials, we should mention bulk Rashba spin splittings in non-centrosymmetric BiTeI (Landolt et al. 2012) and in the ferroelectric Rashba semiconductor α-GeTe were the splitting is coupled to the ferroelectricity (Krempaský et al. 2015).

### 1.3.3 Resonant photoemission: Impurity systems

(Ga,Mn)As with Mn concentrations of a few percent is a paradigm diluted magnetic semiconductor (DMS) where the Mn doping induces its FM properties associated with the hole carriers. Essential for physics of (Ga,Mn)As is the exact energy alignment of the Mn $3d$ derived impurity state (IS) and its hybridization with the GaAs host states. With Mn atoms replacing only a few percent of Ga, finding their response between the bulk of GaAs atoms is a "needle in a haystack" problem which solves however by taking into advantage the penetrating ability and chemical specificity of resonant SX-ARPES.

The experiments (Kobayashi et al. 2014) were performed on (Ga,Mn)As thin films embedding Mn atoms with a concentration of 2.5% and capped by an amorphous As layer to reduce oxidation during ex-situ sample transfer (Kobayashi et al. 2012). Fig. 4(*a*) shows the experimental X-ray absorption (XAS) spectrum at the Mn $L_3$-edge, and (*b*) the corresponding SX-ARPES images measured through the resonance. The pre-resonance image is hardly distinguishable from bare GaAs showing the light-hole band. However, tuning $h\nu$ to the first XAS peak at 640 eV, corresponding to the FM substitutional Mn ions, immediately unleashes the Mn $3d$ derived impurity state (IS) just below $E_F$ which injects the FM charge carriers. Furthermore, simultaneous enhancement of the GaAs band identifies its hybridization with the IS. The second XAS peak at 640.5 eV, corresponding to the PM interstitial Mn atoms, leaves only an afterglow of the IS, and further increase of *hv* returns us to bare band structure of the GaAs host. The ferromagnetic IS can therefore be seen only at one single excitation energy, corresponding to X-ray absorption of the FM substitutional Mn ions. In this way, chemical specificity of resonant SX-ARPES allows us to probe impurity states and their hybridization with the host states for impurity systems with atomic concentrations of a few percent.

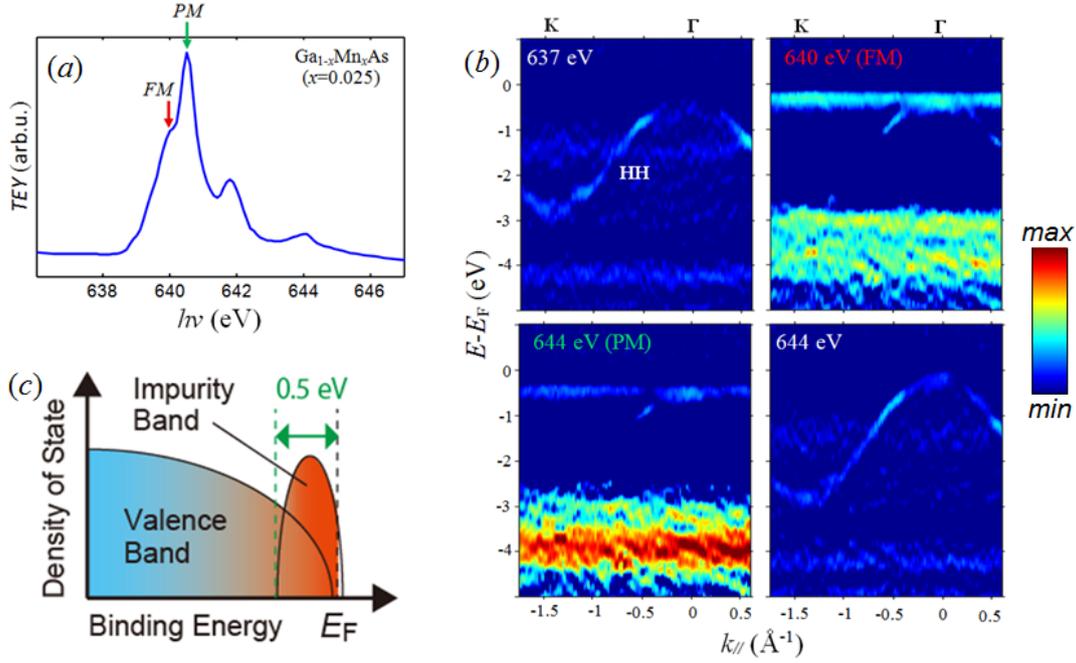

**Fig. 4:** SX-ARPES of the (Ga,Mn)As impurity system: (*a*) Mn $L_3$ XAS spectrum showing the ferromagnetic (FM) and paramagnetic (PM) components; (*b*) ARPES images taken across the resonance. Tuning *hv* onto the FM peak unleashes the Mn impurity state in vicinity of $E_F$ injecting the FM charge carriers; (*c*) Established band diagram. Adapted from Kobayashi et al. (2014).

The band diagram of (Ga,Mn)As emerging from the results of Kobayashi et al. (2014) is sketched in Fig. 6(*c*). It merges the two previous models of (Ga,Mn)As (Ohya et al. 2011; Gray et al. 2012): Whereas the FM charge carriers supplied by the IS located near $E_F$ are characteristic of the double-exchange model, the IS hybridization with the host GaAs is characteristic of the *p-d* exchange model. The impurity character of FM charge carriers in (Ga,Mn)As explains their small mobility, impeding high-speed applications of this DMS. Recently Kobayashi et al. (unpublished) have extended this study to Be-doped InFeAs which is a potential n-type counterpart of GaMnAs for spintronics. In contrast to GaMnAs, the FM charge carriers in InFeAs were found to originate from the dispersive band states at the CB minimum. Their high mobility justifies applications of (In,Fe)As for high-speed spintronics.

An extension of the above study to (In,Mn)As quantum dots buried in GaAs has been reported by Bouravleuv et al. (2016). Of other magnetic impurity systems, we should mention a recent study on the multiferroic compound (Ge,Mn)Te (Krempaský et al. 2016). Mn doping of the host α-GeTe broke the time-reversal protected degeneracy of the Rashba bands in the Γ point. The magnitude of the corresponding Zeeman gap scaled up with Mn concentration, reflecting buildup of ferromagnetic order between the Mn atoms. Another system is V-doped topological insulator $Bi_2Se_3$ where ResPES has found large V weight at $E_F$ (Krieger et al, 2017) whose coexistence with the quantum anomalous Hall (QAH) effect suggests nanoscale separation of this materials into V-rich magnetic and V-deficient non-magnetic phases.

# 2 k-resolved electronic structure of LAO/STO

The interface between LaAlO$_3$ (LAO) and SrTiO$_3$ (STO) is a paradigm example of new functionalities formed at TMO interfaces of TMOs. Although these materials in their bulk form are band insulators with wide band gaps (5.6 eV for LAO and 3.2 eV for STO) their interface spontaneously forms mobile two-dimensional electron system (2DES). Its high electron mobility typical of uncorrelated electron systems co-exists with superconductivity [for entries the reviews Mannhart and Schlom (2010); Hwang et al. (2012)], ferromagnetism (Bannerjee et al. 2013), large magnetoresistance (Caviglia et al. 2010) and other phenomena typical of localized correlated electrons (see Chapter 8). These properties of the 2DEG can be tuned with field effect allowing design of transistor structures with enhanced functionalities (Hosoda et al. 2013; Woltmann et al. 2015).

The interfacial 2DES is confined within a narrow region of a few nanometres on the STO side (Basletić et al. 2008; Dubroka et al. 2008; Son et al. 2009; Delugas et al. 2011; Gariglio et al. 2015) where the mobile electrons populate the $t_{2g}$-derived $d_{xy}$-, $d_{xz}$- and $d_{yz}$-states of Ti ions acquiring reduced valence compared to the bulk Ti$^{4+}$. Breaking of the 3D periodicity at the interface splits the $t_{2g}$ bands into the $d_{xy}$ and two degenerate $d_{xz}/d_{yz}$ bands. Furthermore, the 2DES confinement in the interface quantum well (QW) within a narrow region of a few nanometres on the STO side further splits the $d_{xy}$ and $d_{xz}/d_{yz}$ bands into a ladder of subbands (Breitschaft et al. 2010; Delugas et al. 2011; P.D.C. King et al. 2014; Cancellieri et al. 2014). This complex energy structure based on the correlated 3$d$ orbitals, very different from conventional semiconductor heterostructures described as free particles embedded in the mean-field potential, is the source of a rich and non-trivial phenomenology. Instrumental to explore the underlying band structure of the LAO/STO interface states is SX-ARPES with its virtues of **k**-resolution, enhanced probing depth and chemical specificity. The SX-ARPES results presented below were achieved at the ADRESS beamline of the SLS.

## 2.1 Resonant photoemission

Despite the enhanced probing depth of SX-ARPES, the buried interface 2DES can only be accessed by boosting its photoemission response using ResPES at the Ti $L$-edge. As we will see below, tts elemental and chemical state specificity (Olson 1996; Molodtsov et al. 1997) allows discrimination of the signal coming from the interface Ti$^{3+}$ ions whose valence contrasts them to the Ti$^{4+}$ ones in the STO bulk.

A typical Ti $L$-edge XAS spectrum of the LAO/STO heterostructure measured in total electron yield (TEY) is shown in Fig. 5 (*a*). Almost independent of X-ray polarization, the spectrum shows two leading peaks around 458 and 460.5 eV which correspond to electron excitation from the 2$p_{3/2}$ core levels to 3$d$ $t_{2g}$ and $e_g$ levels, respectively, of the Ti$^{4+}$ ions. The corresponding XAS weight from the Ti$^{3+}$ ions in the region of $hv \sim 459.3$ eV, which as can be seen in the LaTiO$_3$ XAS data (Salluzo et al. 2013; Cancellieri et al. 2013), is hardly seen in the experimental spectrum behind the overwhelming Ti$^{4+}$ contribution. The next pair of peaks in the XAS spectrum corresponds to electron excitation from the 2$p_{3/2}$ core levels (for detailed analysis of XAS spectra of LAO/STO see Chapter 11).

The corresponding map of angle-integrated ResPES intensity in Fig. 5 shows four salient peaks in the VB region. Their energy position reflects the Ti$^{4+}$ peaks in the XAS spectrum, identifying hybridization of the oxygen valence states with Ti$^{4+}$. Most important, at the excitation energies corresponding to the Ti$^{3+}$ regions of the XAS spectrum the ResPES map shows clear signal of the Ti derived 2DEG near $E_F$. This observation arms us with the exact recipe how to experimentally access the 2DES (Koitzsch et al. 2011, Berner et al. 2013, Cancellieri et al. 2013). The corresponding EDCs should be taken away from the stray Ti 2$p$ core level signal excited by second-order radiation from the beamline monochromator. We note that the present ResPES spectra, characteristic of the standard LAO/STO samples, are significantly modified by oxygen deficiency (see Sec. 4.1).

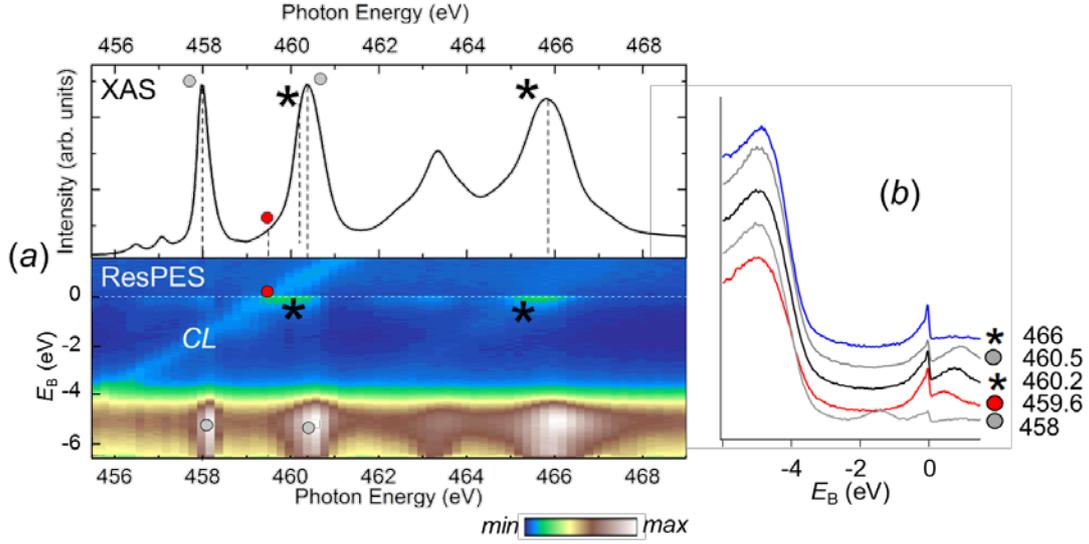

**Fig. 5**. ResPES of standard (oxygen annealed) LAO/STO: (*a*) XAS spectrum; (*b*) Map of angle-integrated ResPES intensity. *CL* marks the stray signal from the Ti 2*p* core level excited by second-order radiation; (*c*) EDCs extracted from (*b*) at selected excitation energies: the 458 and 460.5 eV are $Ti^{4+}$ resonant, 459.6 eV gives the highest signal at $E_F$ but overlaps with the second-order Ti 2*p* signal, and 460.2 and 466 eV are representative of the 2DEG. Adopted from Cancellieri et al. (2013).

Before switching to **k**-resolved band structure of LAO/STO, we should mention a possibility to use ResPES measurement for XPS depth profiling of the 2DES. In the soft-X-ray energy range, measurements at room temperature suppress **k**-selectivity of the photoemission process (so-called XPS regime, see Braun et al. 2013). In this case angle dependence of photoemission intensity $I(\theta)$ reflects attenuation of the 2DES signal in the atomic overlayers. Mathematically, this can be expressed as

$$I(\theta) = G \int_0^\infty R(z) e^{-\frac{z}{\lambda \cos\theta}} dz \qquad (1)$$

where *G* is a geometrical factor, *R*(z) the spatial profile of the 2DES, and the exponent represents the photoelectron attenuation in the overlayers following the Beer-Lambert law (see, for example, Paynter 2009). Cancellieri et al. (2013) have used the expression (1) to fit their experimental $I(\theta)$ data measured at the $L_3$ absorption edge, Fig. 6 (*a*), with *R*(z) approximated as a rectangular shaped function with a thickness *d* (*b*).The fit (*c*) has revealed that the 2DES is localized within d = 1 u.c. at the STO side of the interface.

Such narrow localization of the 2DES seems to be at odds with a number of other works suggesting the 2DES extension by at least a few nm into the STO bulk (Copie et al. 2009; Gariglio et al. 2015). This controversy can be resolved based on later **k**-resolved SX-ARPES measurements (discussed in the next section). Fig. 7 (*a*) demonstrates that for excitation energy at the $L_3$ absorption edge the ResPES intensity is dominated by the $d_{xy}$ band, and this band is indeed localized next to the interface. Moreover, as this region embeds the $Ti^{3+}$ ions with maximal electron occupancy, the present results qualitatively agree with XPS depth profiling of the $Ti^{3+}$ core level in the hard-X-ray energy range performed by Sing et al. (2009). We note that both XPS results were measured at room temperature which can also contribute to their difference to those of Copie et al. (2009) and Gariglio et al. (2015) measured at low temperature. In contrast to the $d_{xy}$ bands, the $d_{yz}$ ones extend much further into the bulk accompanied by gradual transformation of the Ti ions to the $Ti^{4+}$ bulk valence state. These bands actually dominate the ResPES response at the $L_2$ absorption edge, Fig. 7 (*a*). Therefore, the XPS depth profiling would see there much larger extension of the 2DES.

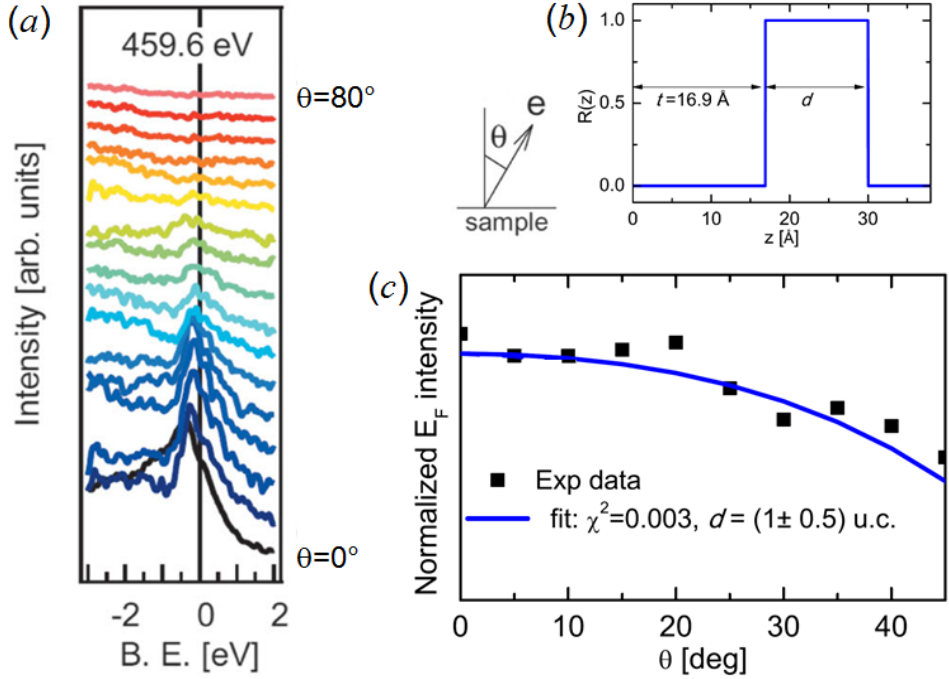

**Fig. 6**. XPS depth profiling of the 2DES: (*a*) Dependence of the ResPES intensity at $hv$ = 460.2 eV and (*b*) its fit within the model of rectangular 2DES distribution (*c*). The fit finds the interfacial $\underline{d}_{xy}$ states within 1 u.c. in STO. Adopted from Cancellieri et al. (2013).

## 2.2 Electronic structure fundamentals: Fermi surface, band structure, orbital character

Band structure and FS are fundamental characteristics of the interfacial 2DES. Their **k**-resolved measurements with SX-ARPES are performed using ResPES (Berner et al. 2013; Cancellieri et al. 2014) with the sample cooled below ~30K in order to suppress the thermal effects reducing the coherent spectral weight (see Sec. 1.1). Measurements of the band dispersions, including the heavy $d_{xz/yz}$-derived bands, require challenging figures of energy resolution better than 40 meV, which for measurements of the FS can be relaxed to ~100 meV.

Typical experimental band structure of the standard LAO/STO samples along the ΓX direction of the BZ measured at two different X-ray polarizations and excitation energies is represented in Fig. 7 (Cancellieri et al. 2016). With the experimental geometry shown in Fig. 1 (*b*), *s*-polarization selects the $d_{xy}$- and $d_{yz}$-bands (Cancellieri et al. 2014) whose wavefunction is antisymmetric relative to the ΓX symmetry line (Damascelli et al. 2003, Strocov et al. 2014-2). The corresponding SX-ARPES data measured at the $L_3$- and $L_2$-resonances is shown in Fig. 7 (*a*) respectively. By comparison with the overlaid $E(\mathbf{k})$ dispersions, calculated in the framework of self-interaction corrected DFT (see Chapter 8), we immediately recognize the heavy $d_{yz}$-band. The $d_{xy}$-subbands are not visible due to vanishing matrix elements, but they manifest themselves as two bright spots where they hybridize with the $d_{yz}$-band. Photoexcitation at the $L_3$-resonance delivers maximal intensity to the $d_{xy}$-bands and $L_2$-resonance to the $d_{yz}$-band, identifying orbital selectivity of resonance photoexcitation. Already at this point, we note strong renormalization of the $d_{yz}$-band compared to the DFT, and waterfalls of spectral intensity extending from the $d_{xy}$-spots. As we will see in Sec. 3.2, these spectral features manifest polaronic nature of the interface charge carriers. *p*-polarization of incident X-rays switches the ARPES response to the Ti $d_{xz}$-wavefunctions symmetric relative to the ΓX line (Cancellieri et al. 2014). The corresponding SX-ARPES data is shown in Fig. 7 (*b*) for the $L_3$- and $L_2$-resonances. We note remnant signal from the $d_{yz}$-band manifesting slight structural distortions at the interface (Zhong and Kelly 2008) relaxing the perfect symmetry.

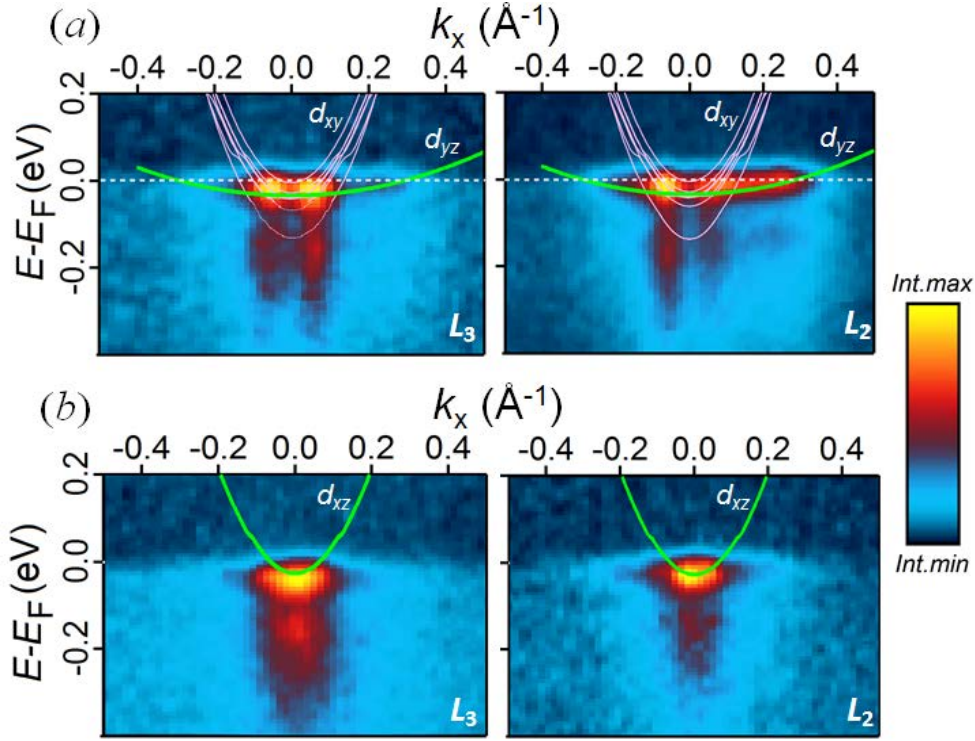

**Fig. 7.** Band structure of standard LAO/STO samples, measured at (*a*) *s*-polarization of X-rays, exposing the $d_{xy}$- and $d_{yz}$-bands, and (*b*) *p*-polarization, exposing the $d_{xz}$-bands. The DFT calculated dispersions are indicated. Renormalization of the experimental bands and tails of their spectral weight extending to lower $E_b$ identify polaronic nature of the 2DES. The $L_3$- and $L_2$-resonance data were collected at $h\nu$ = 460.2 eV and 466 eV, respectively. Adapted from (Cancellieri et al. 2016).

Typical FS of the standard LAO/STO samples collected with circular X-ray polarization at the $L_3$-resonance and extending over four BZs (Cancellieri et al. 2014) is shown in Fig. 8 (*a*). These results are in agreement with the pioneering experiments by Berner et al. (2013). The FS contours in the $\Gamma_0$ point show a circle, originating from the light $d_{xy}$ bands, and two ellipsoids aligned along the $k_x$ and $k_y$ directions, originating from the heavy $d_{yz}$ and $d_{xz}$ bands. The intensity asymmetry relative to the horizontal ΓX line reverses with the X-ray chirality, manifesting a circular dichroism. Interestingly, the apparent FS shapes significantly vary through the four BZs. This effect is caused by different dependence of the $d_{xy}$, $d_{xz}$ and $d_{yz}$ photoemission matrix elements and thus corresponding FS sheets on ($k_x$, $k_y$). Furthermore, in line with linear dichroism of the band structure in Fig. 7, one can use linear X-ray polarization for orbital decomposition of the experimental FS, Fig. 8 (*b*): *s*-polarization enhances the $d_{xy}$ and $d_{xy}$ sheets, and *p*-polarization the $d_{xz}$ one. Another interesting point is that the apparent FS contours are somewhat different between the $L_2$- and $L_3$-resonances (not shown here for brevity) as a consequence of the orbital selectivity of resonant photoemission mentioned above. Evolution of band structure and FS of the LAO/STO interface a function of LAO overlayer thickness was studied by Plumb et al. (2017).

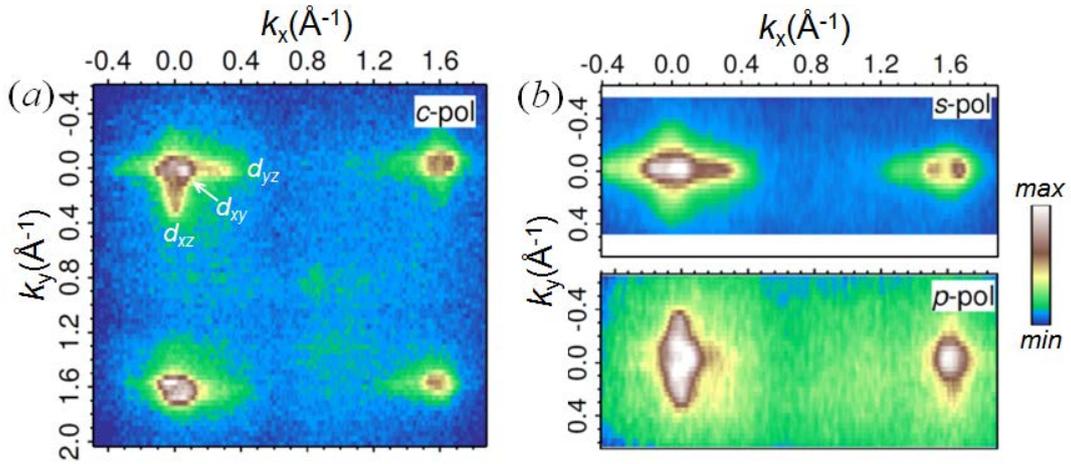

**Fig. 8.** FS of standard LAO/STO samples collected at circular (*a*) and linear (*b*) X-ray polarizations. *s*-polarization enhances the $d_{xy}$ and $d_{xy}$ sheets, and *p*-polarization the $d_{xz}$ one.

### 2.3 Doping effect on the band structure

In Chapters 2-3, it is well explained how the growth conditions affect the electronic properties like conductivity, concentration $n_s$ and mobility of the charge carriers. In particular, $n_s$ changes consistently with the growth temperature (Fête et al. 2015) as well as with oxygen annealing (Cancellieri et al. 2010), although in the latter case the conductivity can transform from 2D to 3D. Cancellieri et al. (2014) have performed SX-ARPES studies of LAO/STO samples with different doping levels in order to correlate the band occupancy and FS shape with $n_s$ measured in transport. Standard doped (SD) samples with $n_s = 6.5 \cdot 10^{13}$ e/cm$^2$ and low doped (LD) samples with $n_s \sim 10^{13}$ el/cm$^2$ were investigated. The ARPES dispersions were compared with theoretical simulations performed with the transport $n_s$ value in order to clarify the corresponding orbital-resolved charge distributions and band occupations (a detailed theoretical analysis of the LAO/STO band structure is provided in Chapter 8).

Theoretical simulations for the SD sample in Fig. 9 (*a*) show that for given $n_s$ most of the charge is embedded in three bands. The two lowest have the planar $d_{xy}$ character with a light mass $m* \approx 0.7 m_0$ ($m_0$ is the free-electron mass) in the ΓX direction. The associated electron charge is confined entirely within the first and second TiO$_2$ layers from the interface. The third occupied band has the $d_{yz}$ character and embeds the charge spread out-of-plane. The band is heavy along $k_x$ ($m* \approx 9 m_0$) and shifted upward in energy by ~70 meV relative to the lowest $d_{xy}$ band. The corresponding simulated FS is consistent with the experiment, Fig. 9 (*b*). Luttinger count of the simulated FS demonstrates that even for the SD samples with their maximal $n_s$ the ideal limit dictated by the polar catastrophe ($n_s = 3.3 \cdot 10^{14}$ e/cm$^2$ corresponding to half an electron per interface u.c.) is not actually achieved.

With decrease of doping for the LD sample, obtained by reduction of the growth temperature (Fête et al. 2015), the experimental FS reduces its Luttinger count as expressed by reduction of the $d_{yz}$-band Fermi vector $k_F$ by ~0.05 Å$^{-1}$, which confirms coherent nature of the interfacial electron transport. The observed trend is consistent with the theoretical simulations in Fig. 9 (*d*) incorporating the reduced $n_s$, where the whole band manifold shifts up in energy. The agreement, however, stays on the qualitative level because the experimental $d_{yz}$ band stays partly occupied (although strongly reduced in its spectral weight), whereas in the calculations the $d_{yz}$ band is entirely above $E_F$. Furthermore, we notice that for both SD and LD samples the experimental bands are systematically lower compared to the calculations. In other words, ARPES captures more charge carriers than detected in the transport measurements. This discrepancy hints on electronic localization and/or phase separation mechanisms at the LAO/STO interface (see Sec. 4.4). Under yet smaller doping the out-of-plane $d_{yz}$ and $d_{xz}$ orbitals will fully depopulate (Lifshitz transition) and the whole mobile charge will reside in the $d_{xy}$ bands close to the interface with concomitant dramatic change in transport properties. Such depopulation can actually be achieved in gating experiments (Chapter 3).

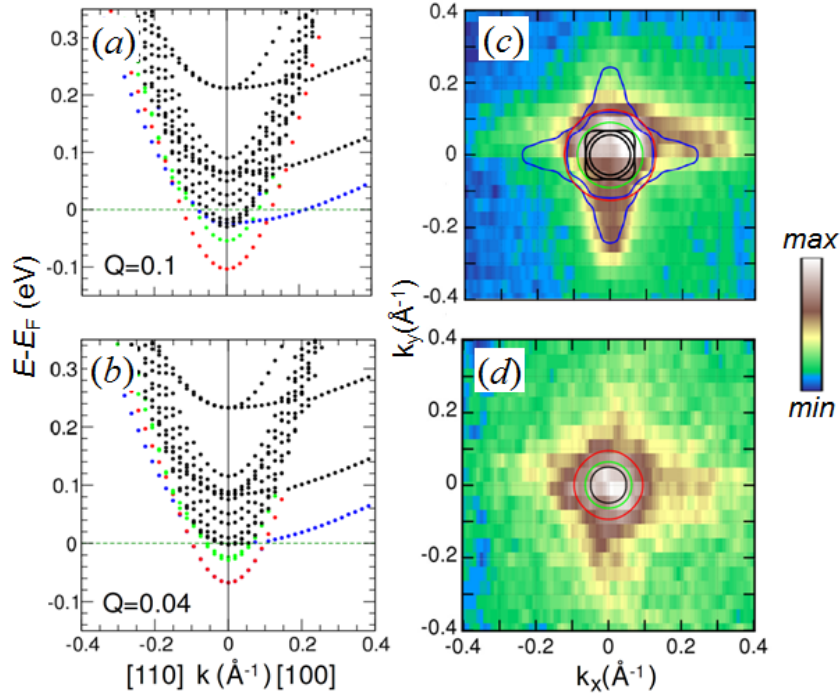

**Fig. 9.** Theoretical band structure and FS (calculated with the transport $n_s$ value) compared with the experimental FS: (*a-b*) Standard and (*c-d*) low doped LAO/STO sample. Luttinger count of the experimental FS follows $n_s$. Adapted from (Cancellieri et al. 2014).

# 3 Electron-phonon interaction and polarons at LAO/STO

We will review now SX-ARPES results that reveal strong electron-phonon interaction (EPI) at the LAO/STO interface. This effect results in polaronic nature of the interfacial charge carriers expressed by large Holstein-type polarons. This phenomenon plays a crucial role in transport properties of LAO/STO. For related physics of bare STO interface see Chapter 4. We start with a recap of fundamentals of the polarons.

## 3.1 Basic concepts of polaron physics

### 3.1.1 Qualitative picture of polarons

The physics of a particle interacting with its environment – a polaron – is a general problem historically started from a Fröhlich model describing an electron moving in a dielectric medium (Fröhlich 1950, Appel 1968) and disturbing neighboring lattice by long-range Coulomb forces. Very soon it was realized that similar phenomenon occurs in molecular crystals where a Holstein model (Holstein 1959) is applied assuming that electron is locally coupled to the host molecule deformations. Nowadays, it is already well known that depending on a "particle" and "environment" and how they interact with each other, the polaron concept applies to extreme diversity of physical phenomena (Alexandrov 2007, Devreese and Alexandrov 2009).

A simple physical picture behind the polaron phenomenon can be imagined as an electron moving in a crystal where ions displace from their equilibrium positions in response to electron-lattice interaction. This electron dragging behind a local lattice distortion – or phonon "cloud" – forms the polaron as composite charge carrier whose properties are significantly different from the bare electron, e.g. the polaron mass $m^*$ is heavier than the bare electron mass $m_0$. Formation of polarons significantly modifies the properties of the system. Apart from changing $m^*$ (Hague et al. 2006, Mishchenko et al. 2000), it modifies optical conductivity (Mishchenko et al. 2003, De Filippis et al. 2006), mobility (Mishchenko et al. 2015), ARPES response (Mishchenko et al. 2000), etc.

We restrict our following discussions to the typical situations when electrons interact with a dispersionless optical phonons of frequency $\omega_0$. In this case the electron-phonon interaction (EPI) can by described by the Hamiltonian

$$H_{int} = \frac{1}{\sqrt{N}} \sum_{\mathbf{kq}} \Gamma(\mathbf{k},\mathbf{q}) \left( c_{\mathbf{k}}^+ c_{\mathbf{k}-\mathbf{q}} b_{\mathbf{q}} + c_{\mathbf{k}-\mathbf{q}}^+ c_{\mathbf{k}} b_q^+ \right)$$

that represents the scattering of electron changing its momentum by $\mathbf{q}$ when it annihilates ($b_{\mathbf{q}}$) or creates ($b_q^+$) a quantum of lattice vibration (phonon) carrying momentum $\mathbf{q}$. Here $c_{\mathbf{k}}^+ / c_{\mathbf{k}}$ are creation/annihilation operators of electron with momentum $\mathbf{k}$.

In general, the interaction vertex $\Gamma(\mathbf{k},\mathbf{q})$, characterizing the EPI dependence on momenta of the particles participating in the scattering event, depends not only on phonon $\mathbf{q}$ but also on electron $\mathbf{k}$. Although the $\mathbf{k}$-dependence leads to a plenty of interesting phenomena (Marchand et al. 2010) it is usually observed either in organic materials (Su et al. 1979, Heeger et al. 1988) or arises in case of magnetic polarons (Liu and Manousakis 1992). Limiting our discussion to lattice polarons in inorganic materials, we consider only a $\mathbf{q}$-dependent vortex $\Gamma(\mathbf{q})$. However, even within this simplification there remains a variety of profoundly different EPI behaviors which are usually divided into long-range and short-range interaction.

### 3.1.2 Long-range vs. short-range electron-phonon interaction

The effect of EPI on physical properties depends on its long- vs. short-range character. The long-range EPI with optical phonons is represented by so-called Fröhlich interaction arising from Coulomb coupling having maximal strength at zero momentum

$$\Gamma(\mathbf{q}) \sim \frac{\sqrt{\alpha}}{q^{(d-1)/2}}. \tag{2}$$

Here $\alpha$ is a dimensionless coupling constant and $d = 2$ or 3 is the dimensionality of the system (Peeters et al. 1986). Square of the potential (2) is the Fourier transform of the long-range three-dimensional Coulomb potential $1/r$. Thus, the EPI (2) implies three-dimensional interaction between the electron and polarization although the electron movement is embedded into $d$-dimensional space. One should note that the Fröhlich interaction does not introduce singularities into physical properties of ground state, because its contribution into a physical quantity, e.g. energy renormalization $\Delta E$, implies the radial momentum integration

$$\Delta E \sim \int_0^\infty \frac{|\Gamma(\mathbf{q})|^2}{\omega_0 + q^2/(2m_0)} q^{d-1} dq$$

whose radial dimensionality-dependent factor $q^{d-1}$ cancels the singularity of $\Gamma(\mathbf{q})$ at zero momentum.

The short-range EPI is represented by so-called Holstein interaction (Holstein 1959) which is momentum independent

$$\Gamma(\mathbf{q}) = g \qquad (3)$$

and arises from different mechanisms of short-range coupling to charge density distortions (Allen 1977, Bilz et al. 1979). Note that the momentum independence (3) implies, in contrast to the long-range one (2), that the interaction is local in the direct space as described by the delta-functional $\delta(\mathbf{r})$. Extending the strict definition (3), a terminology of Holstein-like EPI is often used for the cases where $\Gamma(\mathbf{q})$ is maximal at the Brillouin zone (BZ) boundaries (Kikoin and Mishchenko 1990, Slezak et al. 2006) which is typical of breathing phonon modes.

The EPI for the Holstein model is traditionally characterized by the dimensionless coupling constant

$$\lambda = \frac{2g^2}{W\omega_0} \qquad (4)$$

defined in terms of the bare particle bandwidth $W$. Physically, $\lambda$ is the ratio of two energies, one is the potential well due to lattice deformation and another is kinetic energy helping escape the well. Namely, the former is the polaron binding energy $g^2/\omega_0$ at zero bandwidth $W = 0$ and the latter is half of the bare electron bandwidth $W/2$. In the hypercubic lattice of dimensionality $d$ we have $W = 4dt$ and $m_0 = 1/2t$, where $t$ is the electron hopping constant. If $\Gamma(\mathbf{q})$ is momentum dependent, one can introduce an effective Holstein-like dimensionless coupling constant $\lambda$ by substituting $g \to \Gamma(\mathbf{q})$ and averaging the quantity (4) over the BZ.

The above definition of the dimensionless coupling (4) has several advantages. First, it is compatible with definition of $\lambda$ in theory of metals (Grimvall 1981) if one realizes that $1/W$ is roughly equal to the electronic density of states $N(E_F)$ at the Fermi energy $E_F$. Second, the perturbation theory for mass renormalization in metals gives very simple expression (Grimvall 1981)

$$m^*/m_0 = 1 + \lambda \qquad (5)$$

Note however that in the case of a single Holstein polaron there is an additional coefficient around unity in front of $\lambda$ which depends on the dimensionality, lattice geometry and phonon frequency (Hague et al. 2006, Chandler et al. 2016). Third, an important convenience of the definition (4) is that the critical value $\lambda_c$ of the dimensionless coupling, which separates weak- and strong-coupling regimes, is around unity,

$$\lambda_c \sim 1 \qquad (6)$$

The properties of a polaron with short-range interaction are weakly renormalized at $\lambda < \lambda_c$ and strongly modified at larger $\lambda$ (Hague et al. 2006) resulting in large polaron $m^*$ and strong lattice deformations accompanying such polaron.

The dimensionless coupling constant $\alpha$ of long-range coupling (2) is conventionally defined in terms of the dielectric properties of the crystal, and a rigorous description of this definition is rather lengthy. However, the most important information can be conveyed if we note that for the Fröhlich polaron the expressions (5) and (6) are valid if we substitute

$$\lambda \to \frac{\alpha}{6}$$

Note that in the perturbation expansion for the Fröhlich polaron $m^* = m_0/(1 - \alpha/6) \approx m_0(1 + \alpha/6)$ we use the smallness of $\alpha$. The crossover between the weak- and strong-coupling regimes for the Holstein model can be a rather fast function of $\lambda$, whereas for the Fröhlich model it is always a smooth function of $\alpha$. Properties of the Fröhlich polaron qualitatively change however at $\alpha_c = 6$ (Mishchenko et al. 2000, Mishchenko et al. 2003, De Filippis et al. 2006).

It can be shown that for Holstein coupling the electron self-energy is momentum independent for large enough $E_F$ (Mishchenko et al. 2014). In such cases one can prove (Mahan 1981) that for the Holstein polaron

$$Z_0(m^*/m_0) \approx 1.$$

For the Fröhlich polaron, on the other hand, this relation is violated even for very weak EPI. Instead, the perturbation expressions for the Z-factor $Z_0 = 1 - \alpha/2$ and mass $m^*/m_0 = (1 - \alpha/6)^{-1}$ yield

$$Z_0(m^*/m_0) \approx 1 - \frac{\alpha}{3}.$$

The polaron properties in the strong coupling regime are considerably different from bare particle both for long- (2) and short-range (3) EPI. However, only the latter case is capable of forming so called small-radius polarons when almost the whole lattice deformation is localized at the site of the particle. The reason for such localization is the $\delta(\mathbf{r})$-character of the short-range EPI which can lead to strong on-site deformation leaving almost intact the neighboring sites. Hence, the small-radius polaron can be viewed as a particle with strong deformation at its site which is surrounded by almost unperturbed lattice. In this case the particle hopping to another site is a dramatic event associated with removal of the large lattice deformation at one site and creation of the same strong distortion at the neighbour site. Naturally, coherent movement of a small polaron is easily destroyed either by lattice imperfection or temperature induced lattice potential fluctuation (Alexandrov 2007, Devreese and Alexandrov 2009, Troisi and Orlandi 2006). Such phenomena are impossible in case of long-range EPI because of its $1/r$ long-range character.

One more manifestation of the EPI restricted to systems with short-range couplings is the so called "self-trapping" phenomenon (Rashba 1982, Ueta et al. 1986). Self-trapping means a dramatic transformation of particle properties when system parameters are slightly changed. It occurs at $\lambda \sim \lambda_c$ when a "trapped" particle state with strong lattice deformation around it and the weakly perturbed "free" particle state may have nearly the same energy. Then, the "free" state with small $m^*$ is effectively trapped by an admixture of the "trapped" state with large $m^*$, see Mishchenko et al. (2002) for demonstrations of such mixing.

One has to note that the "self-trapping" phenomenon does not imply real trapping because it does not violate the translational invariance or momentum conservation. The real trapping requires in fact an attractive impurity potential breaking the translational invariance. For polarons, the corresponding lattice deformation cooperates with the attractive potential, making the impurity-assisted trapping much easier compared to bare particles (Shinozuka and Toyozawa 1979, Mishchenko et al. 2009, Burovski et al. 2008, Berciu et al. 2010, Ebrahimnejad and Berciu 2012). For example, localization of a particle in a three dimensional cubic lattice requires local on-site attractive potential $U$ to be larger than its critical value $U_c$ roughly equal to $W/3$ (Koster and Slatter 1954). For the Holstein polarons, however, $U_c$ approaches zero with increase of $\lambda$ (Shinozuka and Toyozawa 1979, Mishchenko et al. 2009, Burovski et al. 2008, Berciu et al. 2010, Ebrahimnejad and Berciu 2012).

### 3.1.3 Spectral function of polarons and its fingerprint in ARPES

The polarons are characterized by the electronic spectral function $A(\mathbf{k},\omega)$ that can be expressed as

$$A(\omega,\mathbf{k}) = Z_0 \delta(\omega - E(\mathbf{k})) + A_{SB}(\omega,\mathbf{k})$$

Here $\delta$ is the Dirac delta-function, $Z_0$ the quasiparticle (QP) weight or Z-factor, $E(\mathbf{k})$ the renormalized QP energy, and $A_{SB}(\mathbf{k},\omega)$ a high energy tail containing phonon sidebands, i.e. states where removal of electron is accompanied by excitation of one or more optical phonons. The normalization of $A(\mathbf{k},\omega)$ to unity as $\int_{-\infty}^{\infty} A(\omega,\mathbf{k})d\omega = 1$ allows its probabilistic interpretation (Mahan 1981) where an electron is removed without (or with) phonon emission with probability $Z_0 < 1$ (or $1-Z_0 < 1$).

The shape of the high energy tail $A_{SB}(\omega,\mathbf{k})$ at zero temperature can be roughly represented as a sum of $l$-phonon sidebands

$$A_{SB}(\omega,\mathbf{k}) = \sum_{l=1}^{\infty} Q_l(\omega) D(\omega + l\omega_0), \quad (7)$$

where $D(\omega) = N^{-1}\sum_{\mathbf{k}}(E(\mathbf{k})-\omega)$ is the normalized density of QP electronic states with dispersion $E(\mathbf{k})$ and $Q_l(\omega)$ are coefficients depending on the EPI strength, full energy width $\tilde{W}$ of the QP electronic band and its filling [for examples see Alexandrov and Ranninger (1992) and Ranninger and Thibblin (1992)]. Energy of each $l$-th sideband in $A_{SB}(\omega,\mathbf{k})$ is shifted from $E(\mathbf{k})$ by $l\omega_0$ (plus, strictly speaking, a slight energy shift due to electron recoil accompanying excitation of phonon). Each sideband is broadened with $\tilde{W}$, and the individual lines in $A_{SB}(\omega,\mathbf{k})$ are resolved when $\tilde{W} \ll \omega_0$ and merge to a single broad hump when $\tilde{W} \gg \omega_0$. The above simple picture of $A_{SB}(\omega,\mathbf{k})$ formed by sidebands separated by one $\omega_0$ breaks down in the cases of self-trapping (Mishchenko et al. 2002) and impurity-assisted trapping (Mishchenko et al. 2009).

So far we considered a model system including only EPI with optical phonons, where the low energy QP pole of $A(\mathbf{k},\omega)$ is a $\delta$-function (Mishchenko et al. 2000). A realistic system contains many other interactions, e.g. electron-electron coupling, interaction with acoustic phonons and impurities. These interactions smear the infinitely high and narrow $\delta$-function, but the remnant peak stays sharp in comparison with the tail of phonon sidebands. As a result, the whole $A(\mathbf{k},\omega)$ acquires a typical peak-dip-hump (PDH) structure (Fig. 10) where a relatively narrow QP peak is followed by broad hump at higher binding energy.

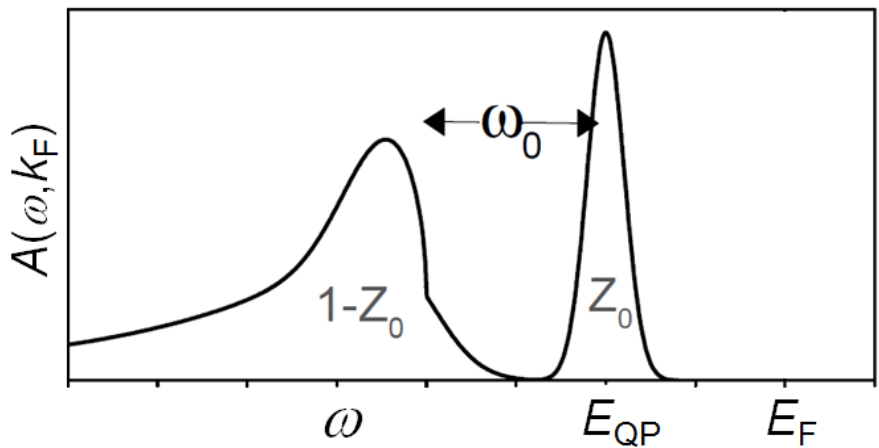

**Fig. 10.** Sketch of the polaron spectral function $A(\omega,\mathbf{k})$ at low electron excitation energy $E(\mathbf{k}) - E_F$. The characteristic peak-dip-hump (PDH) spectral shape is formed by the quasiparticle (QP) peak followed by the phonon sideband tail corresponding to a sequence of optical phonons with frequency $\omega_0$.

The above shape of $A_{SB}(\mathbf{k},\omega)$ is achieved only at low electron excitation energy $E(\mathbf{k}) - E_F \ll \omega_0$. This is however not the case when $E(\mathbf{k})$ is far from $E_F$ (Mishchenko et al. 2000,

Goodvin and Berciu 2010, Hohenadler at al. 2005, Gunnarsson and Rosch 2008). At excitation energy $\gg \omega_0$ the QP dispersion $E(\mathbf{k})$ undresses of the phonon cloud and becomes the bare $\varepsilon(\mathbf{k})$ dispersion. The crossover between these two regimes is characterized by the electron dispersion kink [for entries see Gunnarsson and Rosch (2008); Mishchenko et al. (2011); Lanzara et al. (2001); Cuk et al. (2004); Devereaux et al. (2004)]. A typical picture of this crossover simulated for high-$T_c$ cuprates (Mishchenko et al. 2011) is illustrated in Fig. 11.

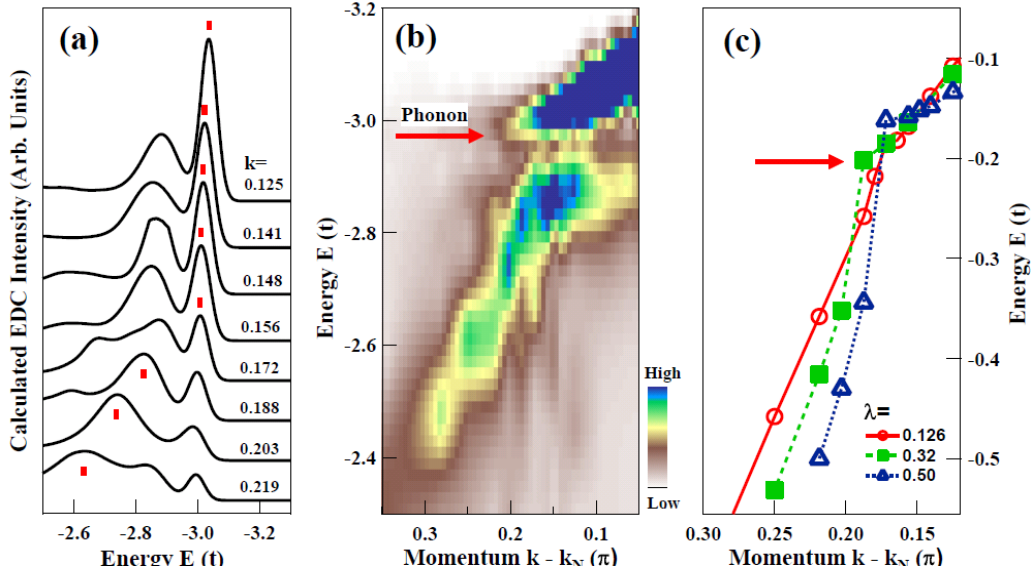

**Fig. 11.** (a) Typical $A(\omega,\mathbf{k})$ of high-$T_c$ cuprates simulated within the $t$-$J$-Holstein model for $\lambda$=0.5 with $\mathbf{k}$ varying along the nodal $(\pi/2,\pi/2) \rightarrow (0, 0)$ direction. The $A(\omega,\mathbf{k})$ maxima are marked; (b) Intensity plot for the same parameters; (c) Momentum dependence of the $A(\omega,\mathbf{k})$ maxima for $\lambda$=0.126 (circles), $\lambda$=0.32 (squares), and $\lambda$=0.5 (triangles). The red arrow in (b) and (c) indicates the optical phonon energy. All energies are in units of $t$ = 0.4 eV measured from the vacuum state of the $t$-$J$ model (a,b) and from the top of the polaron band. Reproduced from Mishchenko et al. 2011.

Naively, the phonon structure of the hump is often imagined similarly to the simplest situation of vibrational excitations in molecules such as gaseous hydrogen (Sawatzky 1989) where the QP peak is followed by a series of sharp lines at multiples of $\omega_0$. This case corresponds to the situation of the independent oscillator model (IOM) (Mahan 1981) whose spectral function at zero temperature $T$ = 0 is described by equidistant lines with weights given by Poisson distribution

$$A_{SB}(\omega) = e^{-g} \sum_{l=0}^{\infty} \frac{g^l}{l!} \delta(\omega - l\omega_0), \qquad (8)$$

where the $l$ = 0 term is the QP peak and $l$ > 0 ones are satellite lines shifted by $l$ energies $\omega_0$. However, this Poisson distribution limit of the general $A_{SB}(\mathbf{k},\omega)$ form (7) can hardly be observed in condensed matter physics because the IOM is valid, first of all, only if the full electronic bandwidth $\widetilde{W}$ is much smaller than $\omega_0$ which is a quite rare case for realistic systems. Furthermore, one needs very low temperature $T \ll \omega_0$ to see the purely Poisson shape of the hump. Indeed, for finite temperature $T$ the IOM spectral function becomes

$$A_{SB}(\omega) = e^{-g(2N+1)} \sum_{l=0}^{\infty} I_l \left\{ 2g[N(N+1)]^{1/2} \right\} e^{\left(\frac{\beta\omega_0}{2}\right)} \delta(\omega - l\omega_0),$$

where $N = \left(e^{-\frac{\beta\omega_0}{2}} - 1\right)^{-1}$ is the Bose distribution function, $I_l$ the modified Bessel function, and $\beta = 1/T$ (Mahan 1981). This function reduces to the Poisson distribution (8) only for $T/\omega_0 \to 0$ and otherwise has no resemblance of it. Note that in case of finite $T$ there appear satellites with negative $l$ values (negative energy loss). The conclusion is that the Poisson distribution of the phonon sidebands can therefore only be awaited when all of the following conditions apply: (i) dispersionless optical phonons $\omega_0$; (ii) excitation energy in vicinity of $E_F$ so that $E(\mathbf{k}) - E_F \ll \omega_0$; (iii) isolated molecule vibrations with vanishing electronic bandwidth $\tilde{W} \ll \omega_0$; (iv) very low temperatures $T \ll \omega_0$. The whole list of these assumptions hardly applies to most of the cases in condensed matter physics.

Although the polarons dramatically influence physical properties of solid-state systems, retrieval of fundamental characteristics of the underlying EPI from various spectroscopic data is not straightforward. For example, $m^*$ can in principle be determined from measurements of optical conductivity $\sigma(\omega)$ using the sum rule (Devreese et al. 1977, Devreese et al. 2010)

$$\frac{1}{m_0} - \frac{1}{m^*} = \frac{2}{\pi e^2 n_0} \int_{\omega_0}^{\omega_m} \sigma(\omega) d\omega,$$

where $e$ is the electron charge, $n_0$ the carriers concentration, and $\omega_m$ the maximal frequency which can be attributed to intra-band conductivity. However, the accuracy of such estimate of $m^*$ depends on how trustworthy is the estimate of $m_0$ provided by band structure calculations.

The unique experimental probe which can capture the polaron physics is ARPES which directly measures the (matrix element weighted) electronic spectral function $A(\mathbf{k},\omega)$ most fully characterizing the EPI. Continuous progress of the ARPES technique reveals more and more cases where the polaronic $A(\mathbf{k},\omega)$ is observed. In particular, the PDH structure was observed, for example, (i) from localized lattice polarons in insulating sodium tungsten bronze $Na_xWO_3$ (Raj et al. 2006); (ii) from lattice polaron states near the dispersion kink in the electron-doped $Nd_{1.85}Ce_{0.15}CuO_4$ high temperature superconductor (Liu et al. 2012); (iii) from lattice polarons at the Ti(100) surface (Moser et al. 2013); (iv) from spin-orbital polaron (Chaloupka and Khaliullin 2007) in misfit cobaltate $[Bi_2Ba_2O_4][CoO_2\$]_2$ (Nicolaou et al. 2010). Lattice polarons at bare $SrTiO_3$ surface are considered in Chapter 4, and those at the $LaAlO_3/SrTiO_3$ interface below in Sec. 3.2. Enhancement of EPI due to disorder was observed by Nie et al. (2015).

Thorough theoretical studies of the ARPES results for high temperature superconductors reveal considerable EPI in these materials (Mishchenko 2009-1, Mishchenko 2009-2). For example, the extended t-J model (Zhang and Rice 1988, Belinicher et al. 1996-1, Belinicher et al. 1996-2), widely used for underdoped materials, well describes (Xiang and Wheatley 1996) the experimental ARPES dispersion (Wells et al. 1995) of the low energy peak in underdoped cuprates. However, while the unbiased Monte Carlo studies of the t-J model predict narrow width of this peak near the band bottom (Mishchenko et al. 2001), the experiment finds the peak width larger than the whole dispersion bandwidth. Inclusion into these theoretical models of EPI extends correct predictions for the underdoped cuprates from the ARPES dispersion (Mishchenko and Nagaosa 2004) to linewidth (Mishchenko and Nagaosa 2004), as well as to temperature dependence of the ARPES spectra (Cataudella et al. 2007-1), kink phenomena (Mishchenko et al. 2011) and optical conductivity (Mishchenko et al. 2008).

## 3.2 Polaronic nature of the LAO/STO charge carriers

### 3.2.1 Low temperature limit

We will analyze the low-temperature SX-ARPES data in Fig. 12 in terms of the one-electron spectral function $A(\omega,\mathbf{k})$, closely following Cancellieri et al. (2016). Fig. 12 reproduces the above *s*-polarization data at the $L_3$- (*a*) and $L_2$- (*d*) resonances together with the corresponding $A(\omega,\mathbf{k})$ profiles for the $d_{xy}$-band along the waterfalls and for the $d_{yz}$-band (*c* and *f*, respectively). Remarkably, the experimental $A(\omega,\mathbf{k})$ reveals a pronounced peak-dip-hump (PDH) structure, where the peak reflects the QP and hump at lower energy its coupling to bosonic modes. Such modes are phonons coupling to electron excitations and forming a polaron (see Sec. 3) which is an electron (hole) dragging behind a local lattice distortion formed in response to strong EPI. The concomitantly increasing $m^*$ of such composite charge carriers fundamentally limits their mobility $\mu \propto 1/m^*$ beyond the incoherent scattering processes. Therefore, the charge carriers at the LAO/STO interface have *polaronic* nature fundamentally limiting the 2DES mobility $\mu_{2DES}$. The experimental $d_{yz}$ dispersion shows the effective mass renormalization $m^*/m^0$ ~2.5 relative to the DFT mass $m^0$. Moreover, clear dispersion of the hump tracking the $d_{yz}$-band, seen in the second derivative plot of Fig. 12 (*e*), identifies a *large polaron* associated with long-range lattice distortions. The hump apex, located at ~118 meV below the QP peak, identifies the main coupling phonon frequency $\omega_0'$. Based on DFT calculations of phonon modes (Aschauer and Spaldin 2014) for bulk cubic STO shown in Fig. 13 (*a*), this phonon can be associated with the hard longitudinal optical mode LO3 at ~100 meV having the largest coupling constant λ among all LO modes (Verbista et al. 1992; Meevasana et al. 2010). This mode is a breathing distortion of the octahedral cage around a Ti site (*b*).

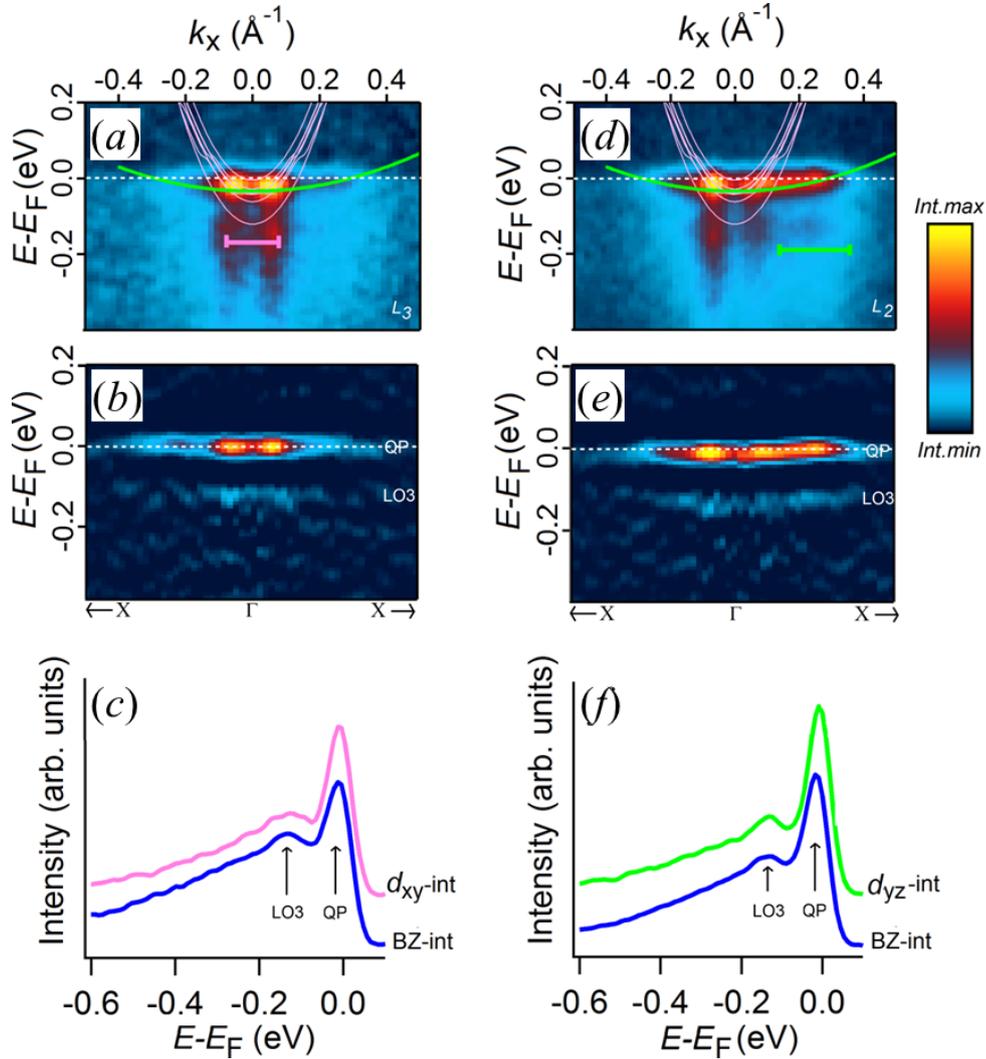

**Fig. 12**. SX-ARPES derived dispersions and $A(\omega,\mathbf{k})$ of the LAO/STO interface states: (*a,d*) ARPES images at the $L_3$ (*a*) and $L_2$ (*d*) resonances with theoretical $d_{yz}$ (green) and $d_{xy}$ (pink) bands. The panels (*b,e*) are their representation in (normalized) second derivative $-d^2I/dE^2>0$; (*c,f*) $A(\omega,\mathbf{k})$ reflecting the $d_{xy}$ and $d_{yz}$ bands integrated over the **k**-regions marked in (*a,c*). The characteristic PDH structure manifests polaronic nature of the interface charge carriers formed by the LO3 phonon, and the hump dispersion identifies a large polaron. Adapted from (Cancellieri *et al.* 2016).

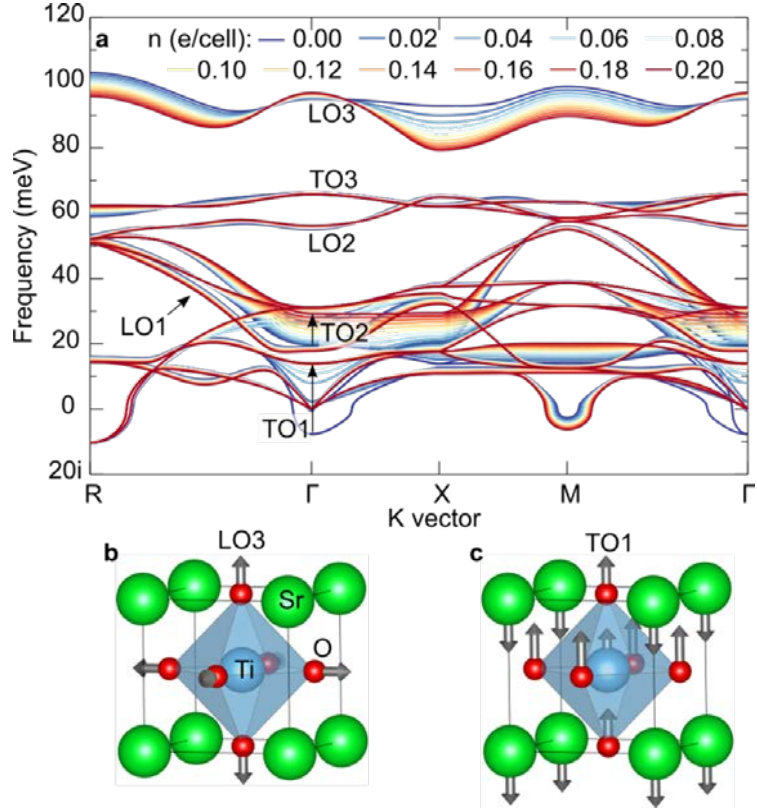

**Fig. 13.** (*a*) DFT calculated phonon modes in doped bulk STO at various electron doping levels $n_v$ (our LAO/STO case corresponds to $n_v$ ~ 0.12); (*b* and *c*) Atomic displacements associated with the breathing LO3 mode and polar TO1 one, respectively (Cancellieri *et al.* 2016).

Optical studies on bulk STO (van Mechelen et al. 2010) have not only confirmed the involvement of the LO3 phonon in EPI but also found the corresponding effective mass renormalization $m^*/m^0$ ~3.0 close to our value. The same phonon was observed by Raman (Cardona 1965) and neutron spectroscopy (Choudhury et al. 2008) as well as by VUV-ARPES at the bare STO(001) surface (Chang et al. 2010; Chen et al. 2015; Wang et al. 2016) with the strength of the polaronic structure depending on $n_s$ (see Chapter 4). The LO3 energy ~100 meV found in all these experiments perfectly matches the DFT calculations, Fig, 13 (*a*). At the LAO/STO interface however the LO3 frequency is somewhat different which can be attributed to coupling of the STO phonon modes to the LAO overlayer. Interestingly, a recent tunnelling study of oxygen isotopic effects at the LAO/STO interface (Boschker et al. 2015) has again found the LO3 phonon (LO4 in the strict tetragonal nomenclature of this work) at ~100 meV. The difference may indicate the fact that whereas the SX-ARPES signal is formed only by coherent electrons coming from the interface region, the tunneling signal has a large contribution of incoherent electrons coming deep from the STO bulk. Nie et al. (2015) have extended VUV-ARPES studies of polaronic effects to the quasi-2D material $Sr_2LaTiO_4$ from the STO-related Ruddlesden-Popper series $Sr_{n+1}Ti_nO_{3n+1}$. La-doping of this material to $Sr_{2-x}La_xTiO_4$ was found to set up an interesting interplay between EPI and disorder, reproduced by model theoretical analysis.

As we have seen above, the polaronic $A(\omega,\mathbf{k})$ can be expressed as $A(\omega,\mathbf{k}) = Z_0 \delta[\omega - E(\mathbf{k})] + A_{SB}(\omega,\mathbf{k})$, where the first term represents the QP peak with dispersion $E(\mathbf{k})$ and the second one the polaronic hump. The experimental $A(\omega,\mathbf{k})$ in Fig. 12 (*b,d*) shows integral weight of the QP peak in the total spectral intensity $Z_0 \sim 0.4$ for both $d_{xy}$- and $d_{yz}$-bands [although the $d_{xy}$-$A(\omega,\mathbf{k})$ shows slightly larger energy broadening resulting from the localization of these states near the defect-reach interface; this spectroscopic observation is consistent with larger defect scattering of the $d_{xy}$ charge carriers found in Shubnikov–de Haas experiments (Fête et al. 2014)]. Crucial for further analysis is the theoretical LO3 phonon dispersions in Fig. 13 (*a*). The LO3 frequency responds to the $n_s$ variations mostly in the large-**q** region. This identifies a short-range character of the corresponding EPI described by the *Holstein-type polaron* (see Sec. 3.1.2). With moderate EPI strength, the Holstein-type EPI forms large polarons, similarly to the Fröhlich polarons driven by long-range EPI (Cataudella et al. 2007-2). The short-range EPI constant $\lambda \sim 1-Z_O$ is in our case ~0.6, which is too weak for the polaronic self-trapping (Millis 1998) but can assist charge trapping on $V_O$s or shallow defects (Mishchenko et al. 2009; Nie et al. 2015) contributing to the polar field compensation in LAO (Hao et al. 2015). Importantly, the QP mass renormalization for the Holstein polarons directly relates to the QP spectral weight as $m^*/m^0 = 1/Z_O$. In our case this relation returns $m^*/m^0 = 2.5$ exhausting the experimental band renormalization in Fig. 12 (*a,c*). This fact leaves not much room for the electron correlation effects in the interfacial 2DES, which is consistent with the tunneling data of Breitschaft et al. (2010). We note also that the significant occupied electronic bandwidth and LO3 dispersion result in significant deviations of the experimental $A_{SB}(\omega,\mathbf{k})$ from the ideal Poisson distribution (see Sec. 3.1.3).

### 3.2.2 Temperature dependence

An intriguing peculiarity of the LAO/STO interface is a dramatic reduction of $\mu_{2DES}$ with temperature *T* (Cariglio et al. 2009) (see Chapter 3). We address this puzzle with a series of *T*-dependent $L_3$ angle-integrated spectra, Fig. 14 (*a*), because analysis of angle-integrated spectra is immune to relaxation of **k**-selectivity with *T* (Braun et al. 2013). The QP peak reduces with increase of *T*, and towards 200K completely dissolves in the phonon hump of $A(\omega,\mathbf{k})$. The corresponding decrease of the QP weight $Z_O$, Fig. 14 (*c*), increases $m^*$ and thus reduces the charge carriers mobility. This microscopic mechanism fully explains the puzzling 2DES mobility reduction observed in Hall effect, Fig. 14 (*b*). In line with our results, optical studies of the bulk STO (van Mechelen et al. 2010) and LAO/STO interface (Dubroka et al. 2008) have also found reduction of the Drude weight with *T*.

The phonon mode behind the observed *T*-dependence can be identified by fitting of the experimental $Z_0(T)$ in Fig. 14 (*c*) with the independent boson model (Mahan 1981) $Z_0(T) = e^{-2g(2N+1)} I_0[2g(2N+1)]$, where $N = (e^{\omega_0/T} - 1)^{-1}$ is the Bose filling factor describing population of the phonon modes as a function of their frequency $\omega_0$ and *T*. The fit identifies the frequency $\omega_0$'' of the soft phonon dominating the EPI as ~18 meV in the low-*T* range and ~14 meV in the high-*T* range. This crossover can be linked to the second-order tetragonal to cubic phase transition in STO at 105K (Wang and Gupta 1973). The theoretical phonon modes in Fig. 13 (*a*) suggest that the observed phonon is the polar TO1 one in bulk STO, Fig. 13 (*c*), sensitive to this phase transition and associated with long-range EPI. The corresponding coupling constant estimated as $\alpha \sim 2$ indicates moderate EPI consistent with the large polaron scenario. The TO1 phonon was also observed in kinks of ARPES dispersions at bare STO(100) (Meevasana et al. 2010) and its hardening due to presence of the 2DES by of the 2DES by THz reflectivity spectroscopy (Nucara et al. 2016). This mode is inherent to the nearly ferroelectric nature of STO, being associated with its huge $\epsilon_{STO}$.

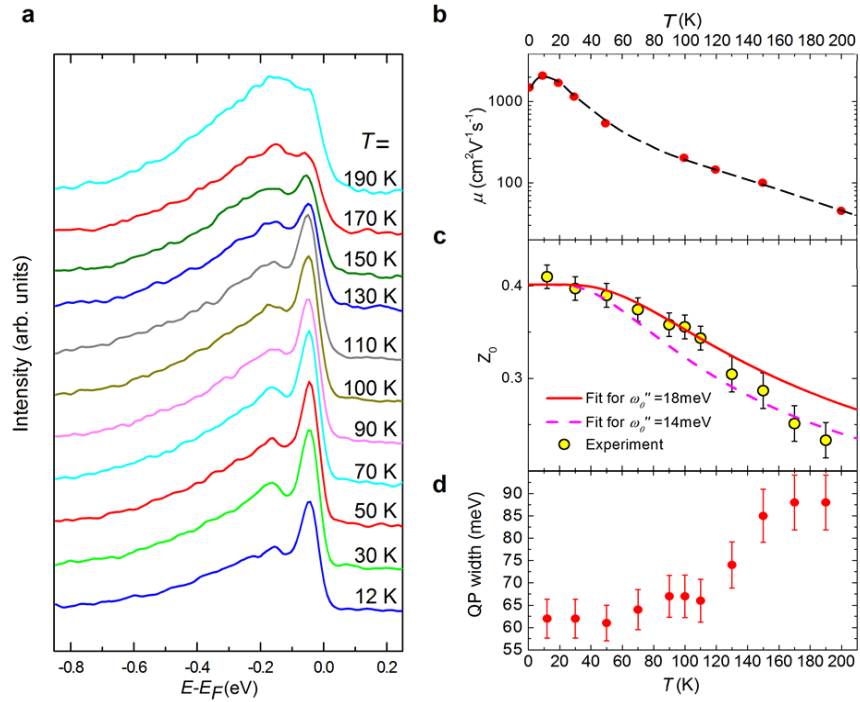

**Fig. 14.** $T$-dependent polaronic effects at LAO/STO. (*a*) $L_3$-resonance angle-integrated spectra. The QP peak dissolving into the hump towards ~190K explains (*b*) the mobility drop with $T$; (*c*) QP spectral weight $Z_0(T)$ fitted by the independent boson model, identifies the soft TO1 phonon, Fig. 13 (*c*), sensitive to the phase transition in STO; (*d*) Corresponding QP energy width (Cancellieri et al. 2015).

The LAO/STO interfacial charge carriers have therefore polaronic nature involving at least two phonons with different energy and thermal activity. The LAO/STO superconductivity, if driven by a phonon mechanism (Boschker et al. 2015), can be related to the discovered polaronic state. Whereas the hard LO3 phonon energy much exceeds the energy scale of the superconducting transition at 0.3K, involved in the electron pairing may be the soft TO1 phonon. The polaronic activity is actually typical of TMO perovskites, reflecting their highly ionic character and easy structural transformations (Wang and Gupta 1973). For further details of polaronic physics at the LAO/STO interface the reader is referred to (Cancellieri et al. 2016).

# 4 Oxygen vacancies at LAO/STO

## 4.1 Signatures of oxygen vacancies in photoemission

Oxygen (ox-) deficiency dramatically affects electronic and magnetic properties of the TMO systems. In particular, transport measurements evidence that the ox-deficiency dramatically increases both concentration and mobility of the interfacial 2DES (Herranz et al. 2007, Cancellieri et al. 2010). For all bulk, surface and interface STO-based systems, each oxygen vacancy ($V_O$) releases two electrons. In general, one of these electrons injected into the 2DES, whereas another stays near the Ti ion (Hao et al. 2015) to form there a localized state. Having a binding energy around 1.2 eV within the band gap of bulk STO, these states are commonly referred to as the in-gap states. Apart from the ARPES results discussed below, the coexistence of mobile and oxygen-derived localized electron states at the LAO/STO interface has been suggested by scanning tunneling spectroscopy (Breitschaft et al. 2010; Ristic et al. 2011), resonant inelastic X-ray scattering (Zhou et al. 2011), O $K$-edge XAS (Palina et al. 2016), etc.

In contrast to the stoichiometric STO where the Ti 3$d$-$e_g$ states are empty, the V$_O$s modify the local covalent bonding in its proximity and cause an orbital reconstruction where the $e_g$ states shift down to become the in-gap states (Pavlenko et al. 2012; Pavlenko et al. 2013; Lin and Demkov 2013). Tight electron localization in this state causes notable correlation effects (Lin and Demkov 2013; Lechermann et al. 2014; Lechermann et al. 2016) which necessitate the use of many-body theoretical methods like the DMFT for their analysis (see Chapter 8). Furthermore, electrons the in-gap states can hop to adjacent lattice sites under thermal or electric field activation (Wang et al. 2011; Lei et al. 2014; Veal et al. 2016) and may therefore be viewed as small polarons (Janotti et al. 2014; Hao et al. 2015) associated with a sizable distortion of the surrounding lattice. Finally, spins of the in-gap states can stabilize the ferromagnetic order (Sec. 4.3). Therefore, the VOs in the STO-based materials give rise to a duality of their electron system (Lechermann et al. 2016; Altmeyer et al. 2016) where delocalized 2DES quasiparticles (which are $t_{2g}$ derived, weakly correlated, non-magnetic and forming large polarons, see Sec. 3.2) co-exist with localized in-gap states ($e_g$ derived, more correlated, magnetic and forming small polarons).

Oxygen (ox-) deficient LAO/STO samples are usually grown at reduced O$_2$ pressure about $10^{-5}$ mBar without the standard post-annealing procedure. Typical ResPES response of these samples is shown in Fig. 15. Whereas the XAS spectrum ($a$) is hardly distinguishable from the annealed samples, Fig. 5, the ResPES intensity ($b$) immediately reveals the V$_O$ derived in-gap states around $E_B$ = 1.2 eV. The $e_g$ character of these states manifests itself in advance of their resonant $h\nu$ compared to the $t_{2g}$-derived 2DES. Interestingly, the in-gap peak slightly displaces in $E_B$ as a function of excitation energy. This effect can trace back either to a multiplet structure of the in-gap states caused by electron correlations, or to different atomic configurations of the V$_O$s.

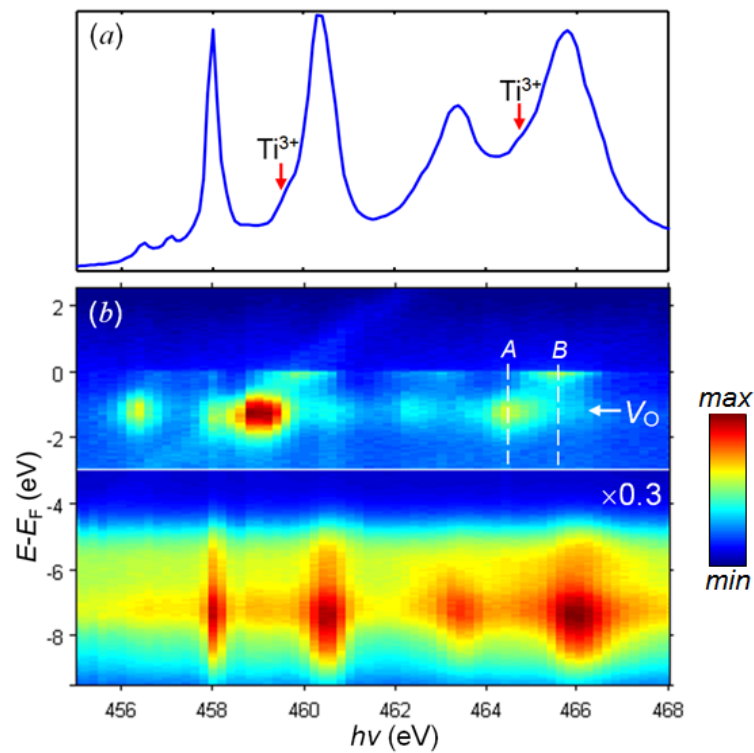

**Fig. 15**. ResPES of ox-deficient LAO/STO: ($a$) XAS spectrum; ($b$) Map of angle-integrated ResPES intensity. The in-gap states resonating around $E_B$ = 1.2 eV are characteristic of the V$_O$s. $A$ and $B$ mark the V$_O$s and 2DES resonance regions, respectively, used in Fig. 17 for the phase separation analysis.

Prototypic of the TMO interfaces are their bare surfaces. VUV-ARPES experiments on TiO$_2$, STO and BaTiO$_3$ surfaces are reviewed in Chapter 4. Oxygen deficiency and thus conductivity of these surfaces can be achieved by their exposure to photons with energy larger than ~38 eV. In this case the formation of VOs involves excitation of Ti 3$p$ core holes which are filled via an interatomic Auger process by electrons from the neighboring O 2$p$ orbitals (McKeown Walker et al. 2015). An alternative method to create V$_O$s is deposition of a thin layer of Al (Rödel et al. 2016) or other metals such as Ti, Nb, Pt, Eu and Sr (Posadas et al. 2017). The delocalized $d_{yz}$ states at the ox-deficient STO interfaces and surfaces can extend to 200 Å and more into the STO bulk (Gariglio et al. 2015) that makes these states nearly three-dimensional (Plumb et al. 2014). Well controlled manipulation by V$_O$s to fine tune the electronic and magnetic properties of TMO heterostructures may assist engineering of future oxide electronic devices.

**4.2 Tuning the polaronic effects**

The polaronic reduction of $\mu_{2DES}$ fundamentally limits the application potential of the STO-based heterostructures. This limit can possibly be circumvented through manipulation of V$_O$s. As we have seen above, in general each V$_O$ releases from the neighbour Ti ion two electrons. One of them stays in the impurity state, and another injects into the mobile 2DES to increases the electron screening and thus reduce the EPI strength (Cancellieri et al. 2016).

The effect of the V$_O$s is illustrated by results of SX-ARPES measurements on oxygen deficient LAO/STO samples, Fig. 17, where increase of the V$_O$ concentration gradually reduces the polaronic weight. This trend is consistent with VUV-ARPES studies at bare Ti(100) and STO(100) surfaces (Moser et al. 2013 and Wang et al. 2016) where $n_s$ was varied in a wide range by exposure to VUV photons creating V$_O$s. The EPI changed in these experiments from very strong, resulting in overwhelming polaronic weight compared to the QP peak, to very weak, surviving in a kink structure of electron dispersions in vicinity of $E_F$. We note, on the other hand, that the V$_O$s can in principle assist the EPI due to charge trapping on the associated shallow defects (Mishchenko et al. 2009; Nie et al. 2015) as well as increase the defect scattering rate (Bristowe et al. 2011), both effects counteracting the above increase of $\mu_{2DES}$.

The effect of the V$_O$s is actually beyond the simple doping picture restricted to changing of the band filling within the rigid band shift model. In particular, the V$_O$s increase the spatial extension of the $d_{xz/yz}$ bands into the STO bulk from ~50 Å for oxygen-annealed samples to ~150 Å and more (Delugas et al. 2011; Son et al. 2009; Basletić et al. 2008) resulting in predominantly bulk conductivity and loss of the 2D nature of the LAO/STO interface system. Furthermore, the manipulation of the V$_O$s is complicated by diffusion processes which are hard to precisely control. An interesting example of "defect engineering" is however the $\gamma$-Al$_2$O$_3$/STO interface (Chen et al. 2013) where not only $n_s$ increases but also low-temperature $\mu_{2DES}$ boosts by almost two orders of magnitude compared to LAO/STO. First-principles calculations suggest that in this case the VOs diffuse out of the STO bulk and accumulate at the interfacial monolayer, reducing thereby their concentration and arguably associated defect scattering in the deeper STO layers (Schütz et al. 2017). Further experiments on ox-deficient LAO/STO interfaces will bring better understanding of the role of V$_O$s and ways to optimize $\mu_{2DES}$.

**4.3 Interfacial ferromagnetism**

Ferromagnetism (FM) of TMO interfaces built up from non-magnetic constituents is one their most interesting functionalities (see, for example, Reyren et al. 2008; Hwang et al. 2012]. The origin of this phenomenon is still elusive even for the paradigm LAO/STO interface (Gabay and Triscone 2013). In one of the theories, so-called Hund's rule induced FM (Gabay and Triscone 2013; Bannerjee et al. 2013), alignment of the motion of the $d_{xy}$ electron spins is mediated by motion of the $d_{xz/yz}$ electrons. In this case the $d_{xy}$ spins may rotate from site to site, giving thus rise to an exotic spin spiral state. An alternative scenario (Herranz et al. 2007; Pavlenko *et al.* 2012; Pavlenko et al. 2013; Saluzzo *et al.* 2013) suggests that the interfacial FM is not an intrinsic property of 2DES and links it to the $V_O$s that spin split the 2DES energy levels. The decisive role of the $V_O$s is consistent with the fact that the interfacial FM quenches with annealing in oxygen.

Hints on the origin of FM at the LAO/STO were recently obtained from VUV spin-resolved ARPES (SARPES) experiments on bare STO surface (Santander-Syro et al. 2015). They suggested that two $d_{xy}$ bands possessed a Rashba-like spin splitting characterized by their opposite spin orientation, Fig. 16(*a*), and helical winding around the corresponding FS sheets (*b*). This spin texture was interpreted as a giant Rashba splitting of ~100 meV enhanced by the antiferrodistortive corrugation of the bare STO(100) surface.as compared to the (field effect tunable) Rashba splitting of a few meV observed by magnetotransport for the LAO/STO interface (Caviglia et al. 2010; Zhong et al. 2013). The most striking result of this study, a giant spin splitting of ~90 meV in the otherwise time-inversion protected Γ-point, was interpreted as an evidence of a FM order induced presumably by the $V_O$s. Subsequent theoretical analysis based on DFT slab calculations (Garcia-Castro et al. 2016; Altmeyer et al. 2016) suggested a modified scenario: The Rashba-like spin splitting of the $d_{xy}$ bands stays within a few meV consistent with the magnetotransport results, but magnetic moments of Ti ions spin split these bands with an energy difference of ∼100 meV at the Γ point consistent with the VUV-ARPES findings. While the magnetism tends to suppress the relativistic Rashba interaction effects, their signatures survive in complex spin textures of the 2DES. The giant spin splitting at the STO(100) surface was however not confirmed by another VUV-SARPES study by McKeown Walker et al. (2016). The conflicting results of the two studies can be reconciled based on a theoretical finding (Lechermann et al. 2014; Behrmann and Lechermann 2015) that while single-$V_O$ configurations possess randomly oriented spins, multiple-$V_O$ configurations can build up the FM order. The PLD grown samples used by Santander-Syro et al. (2015) could have larger concentration of $V_O$s (consistent with larger spectral broadening) compared to the cleaved samples used by McKeown Walker et al. (2016) and thus could embed a larger fraction of the FM multi-$V_O$ configurations. Spin-resolved ARPES will in near future be advanced to understand nature of the LAO/STO interfacial FM using ResPES measurements with multichannel spin detector iMott (Strocov et al. 2015) as discussed in Sec. 5.

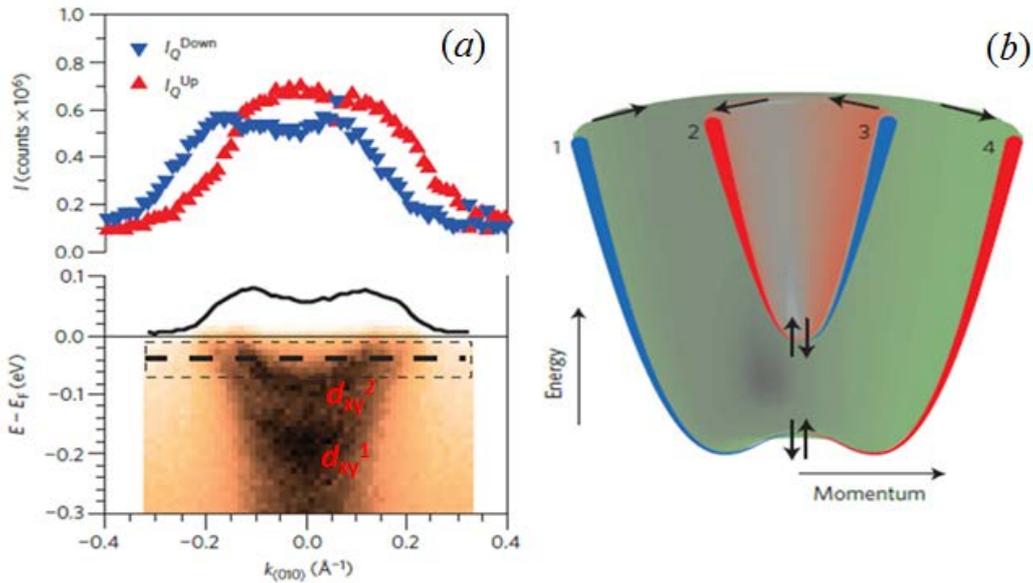

**Fig. 16.** VUV-SARPES of bare STO(100) surface: (*a*) Two $d_{xy}$ bands (bottom) and their spin polarization (top); (*b*) Schematic spin texture. The Zeeman-like splitting of the $d_{xy}$ bands in the Γ-point suggests a FM order connected with $V_O$s. Adapted from (Santander-Syro et al. 2015).

**4.4 Phase separation**

An inherent feature of many TMOs and their interfaces is electronic phase separation (EPS) as spontaneous formation of micro- to nano-scale regions possessing different electronic and magnetic properties. This phenomenon plays crucial role, for example, in colossal magnetoresistance of manganites and stripe order in cuprates [for entries see Dagotto (2005) and Shenoy et al. (2006)]. For the LAO/STO interface, evidence of this phenomenon has been observed by tunneling experiments (Richter et al. 2013; Bucheli et al. 2015), scanning tunneling microscopy/spectroscopy (Breitschaft et al. 2010; Ristic et al. 2011), atomic force microscopy (Bi et al. 2016), magnetoresistance and anomalous Hall effect (Joshua et al. 2013), etc. The EPS has also been identified in a percolative character of the metal-to-superconductor transition with a significant fraction of 2DES resisting superconductivity down to the lowest temperature (Bucheli et al. 2015; Caprara et al. 2014). On the magnetic side, FM puddles embedded in metallic phase have been observed by magnetotransport experiments (Ariando 2011) explaining the intriguing co-existence of the interfacial ferromagnetism and superconductivity. Finally, the EPS explains why $n_s$ observed in transport at the LAO/STO interface always falls short of predictions of the mean-field theories. First ARPES study of the EPS has been performed by Dudy et al. (2016) on bare STO(100) and (111) surfaces as prototypes of the LAO/STO interface accessible with VUV photons. They have found co-existence of metallic and insulating phases, and suggested that this inhomogeneity was driven by clustering of $V_O$s created under VUV irradiation.

Ox-deficient LAO/STO samples offer a convenient platform for studies of EPS in them with ARPES. For this purpose ox-deficient samples are cooled down to room temperature after the growth to stabilize to $V_O$s, and then ex-situ post-annealed in oxygen at ~500$^\text{o}$C. The post-annealing quenches the $V_O$s, but they gradually recover under X-ray irradiation. This puzzling behavior of the post-annealed LAO/STO samples, contrasting them the standard in-situ annealed samples immune to X-rays, is illustrated in Fig. 17 that shows angle-integrated ResPES intensity near the Ti $L_3$-edge at two $h\nu$ values enhancing the $V_O$ (*a*) and 2DES (*b*) intensity (marked in Fig. 15) which depends on X-ray irradiation time (Strocov et al., unpublished). Scaling up of both $V_O$ and 2DES peaks with irradiation reflects the formation of $V_O$s injecting mobile electrons into the 2DES  The total number of mobile electrons in the system, reflected by the 2DES peak, progressively increases with irradiation. On the other hand, Fig. 17 (*b*) shows the corresponding irradiation dependence of **k**-resolved spectral intensity at $E_F$ whose maxima identify $k_F$. Surprisingly, for both $d_{xy}$ and $d_{yz}$ band the $k_F$ values remain constant, identifying constant Luttinger count of the corresponding FS sheets and thus constant $n_s$ in the system [this result does not change if $k_F$ is determined by the gradient method (Staub et al. 1997)]. The observed discrepancy between the increasing total number of mobile electrons and constant $n_s$ identifies EPS at the interface, where the fraction of the conductive interface areas increases with irradiation while their local electronic structure stays constant (strictly speaking, electronic structure might in this experiment evolve but rapidly saturate already with small irradiation dose below the actual time sampling). Extrapolation of these spectroscopic results to the standard LAO/STO samples suggests significant EPS in them as well, in agreement with the above macroscopic evidences. Moreover, the conductive fraction in the standard samples is smaller compared to the ox-deficient ones: the corresponding ResPES data in Fig. 5 (*a*) shows smaller 2DES to Ti 2*p* signal ratio compared to the ox-deficient samples, Fig. 15 (*b*).

Physics of EPS at the LAO/STO interface is a cooperative interplay of the electronic and $V_O$ systems that is far from complete understanding. On the electronic side, recent theoretical analysis (Scopigno et al. 2016) suggested that the lateral confinement allows 2DES to avoid a thermodynamically unstable state with negative compressibility. This effect is particularly strong in the nearly ferroelectric STO due to its huge dielectric constant $\epsilon_{STO}$ screening the electron repulsion and thus allowing accumulation of large electron densities. Another electronic scenario invoked formation of a Jahn-Teller polaronic phase (Nanda and Satpathy 2011). The electronic EPS can be tuned by external electric field (Scopigno et al. 2016). On the $V_O$ side of the EPS, coming into play in ox-deficient samples, theoretical analysis of interplay between the $t_{2g}$ states of the 2DES with the orbitally reconstructed $e_g$ states derived from the $V_O$s (Pavlenko et al. 2013) suggested a complicated phase diagramm with regions of phase-separated magnetic states. Furthermore, the $V_O$s themselves have a tendency to form clusters, which will then accumulate the 2DES (Pavlenko *et al.* 2012; Mohanta and Taraphder 2014). Rationalizing of this intricate and diverse physics of the EPS requires further theoretical and experimental effort combining various area sensitive and local probe spectroscopies.

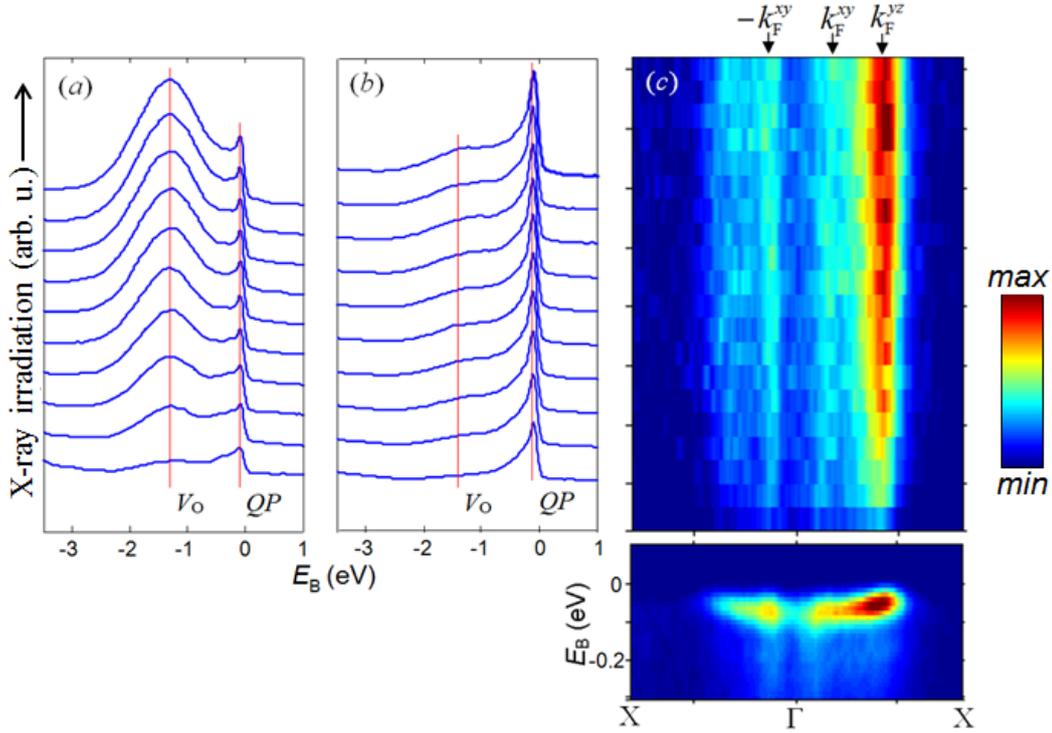

**Fig. 17**. EPS in post-annealed LAO/STO evidenced under X-ray irradiation: (*a,b*) Angle-integrated ResPES intensity for *hν* = 464.6 eV (*a*) and 466.2 eV (*b*) enhancing the $V_O$ and 2DES intensity, respectively (*A* and *B* in Fig.15); (*c*) Saturated *E*(**k**) image (*bottom*) and **k**-resolved intensity at $E_F$ (*top*). The increasing total number of mobile electrons expressed by the 2DES peak (*b*) juxtaposed with constant $n_s$ expressed by the $k_F$ values (*c*) identify the EPS (Strocov et al., unpublished).

# 5 Prospects

SX-ARPES is a method of choice to explore a wealth of intriguing physics in the extremely reach functionality of TMO interfaces. Below we briefly sketch routes to apply SX-ARPES for investigation of the interfacial magnetism and field effect.

*Interfacial spin structure*. Ferromagnetism (FM) of TMO interfaces built up from non-magnetic constituents is one their most interesting functionalities. Understanding of this phenomenon will open new avenues for oxide spintronics such as ferromagnetic Josephson junctions [see, for example, (Komissinski et al. 2007)] where both superconducting and FM constituents use the same TMO materials platform. Switching elements realized on such junctions promise power saving of up to 5 orders of magnitude compared to the nowadays semiconductor-based electronics.

Exploration of the TMO interfacial magnetism requires spin-resolved ARPES measurements (see Chapter 1) where the photoelectron spin is detected, in addition to energy and momentum. Small photoexcitation cross-section of valence states in the SX-ARPES energy range and attenuation of photoelectrons from the buried interfaces, on one side, combined with the immense intensity loss of at least 2 orders of magnitude associated with the spin resolution, on another side, make spin-resolved SX-ARPES experiments extremely difficult. We expect however that this challenge will in near future be resolved with advent of the angle and energy multichannel spin detectors such us the iMott (Strocov et al. 2015) delivering an efficiency gain of a few orders of magnitude.

*Field effect.* This cornerstone property of the LAO/STO interface allowing, for example, fabrication of transistor heterostructures with enhanced functionalities (Caviglia et al. 2008; Hosoda et al. 2013; Woltmann et al. 2015). In-operando SX-ARPES investigations of this effect should use the top-gate geometry, because bias applied from the back gate would hardly affects the QW region next to the interface hosting the $d_{xy}$ states (Scopigno et al. 2016). The latter is consistent with very weak back gating effect observed in soft- and hard-X-ray PE depth profiling (Minohara 2014). The top-gate method suffers however from photoelectron attenuation in the top electrode. Ideal for this purpose would be graphene with its monolayer thickness. Recently, transfer of small (tens of μm) graphene flakes on LAO/STO structures has been demonstrated (Aliaj et al. 2016). A method of TOABr-assisted electrochemical delamination (Koefoed et al. 2015) allows transfer of large (on the inch scale) graphene layers.

Complexity of the field effect at the LAO/STO interface goes much beyond the simple band filling picture typical of the conventional semiconductor heterostructures. First, while population of the deep $d_{xy}$ bands will vary monotonously with bias, the $d_{xz/yz}$ bands placed in only ~40 meV from $E_F$ can completely depopulate at certain negative bias (Lifshitz transition). Given different character of the planar and orthogonal d-orbitals, this transition will dramatically change the optical and transport properties of the LAO/STO interface (Cancellieri *et al.* 2014). Furthermore, variation of the LAO/STO interfacial $n_s$ changes the EPI and thus effective mass and mobility of the charge carriers, resulting in non-linear dielectric response. Oxygen deficiency adds another dimension to this complexity. In this case the bias causes electromigration of the $V_O$s (Lei et al. 2014; Veal et al. 2016) again affecting the interfacial $n_s$ and EPI. Furthermore, the electromigration can affect multiple configurations of the $V_O$s with concomitant effect on their electron correlation and magnetic properties, see Sec. 4.3. We speculate that the latter can potentially open ways to multiferroic functionality of the LAO/STO interface, i.e. using electric field to manipulate its magnetic properties.

*Further materials*. The above application of high-resolution SX-ARPES to the LAO/STO interface taken place at the SLS in 2013 (Cancellieri et al. 2014) has been the first time when **k**-resolved electron dispersions for a buried interface were determined. Having demonstrated its spectroscopic potential, SX-ARPES is now expanding to a wide range of interface and heterostructure systems actual for novel electronic and spintronic devices. We can mention, for example, QWs of STO embedded in LAO (Li et al. 2014) where electron confinement between the two interfaces will form electron spectrum different from the sole LAO/STO interface and tunable by the STO thickness. Superstructures of such QWs (Li et al., unpublished) will allow creation of the artificial third dimension allowing fine tuning of physical properties of these systems. Another fascinating system is the $LaAlO_3/EuTiO_3/SrTiO_3$ heterostructure (Stornaiuolo et al. 2016) where the ferromagnetic $EuTiO_3$ layer induces strong spin polarization of the embedded 2DES tunable by gate voltage in field effect transistor geometry. This system will be an obvious candidate for the spin-resolved SX-ARPES. Other interesting systems may be heterostructures of strongly correlated materials like $SrVO_3$ where tunable interplay between electron correlation and confinement effects can deliver conceptually new electronic devices such as Mott transistors (Zhong et al. 2015). SX-ARPES investigations of multiferroic interfaces are reviewed in Chapter 10. For further scientific and technological cases awaiting applications of (spin-resolved) SX-ARPES the reader is referred to a roadmap of novel oxide electronic materials compiled by Lorentz et al. (2016).

# 6 Conclusions

This chapter has reviewed a number of diverse but interconnected scientific fields ranging from spectroscopic abilities of SX-ARPES to basics of polaron physics and to electronic structure of oxide interfaces. Crucial spectroscopic advantages of SX-ARPES for buried interfaces are its enhanced probing depth and chemical specificity achieved with resonant photoexcitation. We have demonstrated that its application to oxide interfaces delivers direct information on the most fundamental aspects of their electronic structure – momentum-resolved spectral function, band structure, Fermi surface. Focusing on the 2DES formed at the paradigm LAO/STO interface, we have demonstrated determination of its $t_{2g}$ derived multi-orbital band structure and Fermi surface directly connected to the transport properties.

EPI is one of the key players in complex physics of oxides. At the LAO/STO interface, strong electron coupling to the hard LO3 breathing phonon mode forms Holstein-type large polarons. They manifesting themselves as pronounced peak-dip-hump structure of the experimental $A(\omega,\mathbf{k})$ where the quasiparticle peak is followed by broad phonon hump. The polaron formation fundamentally reduces mobility of the interface charge carriers. Furthermore, electron coupling to the soft TO1 polar mode results in dramatic reduction of mobility with temperature. Electron correlations in the interfacial 2DES are weaker compared to the EPI.

Oxygen deficiency adds another degree of freedom to the oxide systems. At the LAO/STO interface, the $V_O$s form a dual electron system where delocalized 2DES quasiparticles ($t_{2g}$ derived, weakly correlated and non-magnetic) co-exist with localized in-gap states ($e_g$ derived, strongly correlated and magnetic). Manipulation by $V_O$s allows therefore tuning of the polaronic coupling and thus mobility of the charge carriers through the 2DES density as well as tuning of the interfacial ferromagnetism critically depending on various atomic configurations of the $V_O$s.

Although SX-ARPES has already produced an impressive amount of fundamental knowledge about the oxide interfaces, this new spectroscopic technique is still in its adolescence period. Ahead are applications to a wide range of interfaces and complex heterostructures, spin-resolved SX-ARPES disclosing spin texture of interfaces, in-operando field effect experiments paving the way towards device applications utilizing the reach physics of oxide interfaces.

# 7 Acknowledgments

We thank M.-A. Husanu, M. Kobayashi, L. L. Lev, V. A. Rogalev, U. Aschauer, A. Filippetti, J.-M. Triscone and others for their contribution to the main scientific cases discussed above, and P. R. Willmott, R. Claessen, M. Sing, M. Radović, F. Baumberger, O.S. Barišić and F. Lechermann for sharing fruitful discussions. Parts of this research were supported by the ImPACT Program of the Council for Science, Technology and Innovation (Cabinet Office, Government of Japan).

39C. Woltmann, T. Harada, H. Boschker, V. Srot, P. A. van Aken, H. Klauk and J. Mannhart, Phys. Rev. Applied **4**, 064003 (2015).

J. J. Yeh and I. Lindau, Atomic Data and Nuclear Data Tables 32, 1 (1985); web version available at http://ulisse.elettra.trieste.it/services/elements/WebElements.html

T. Xiang and J. M. Wheatley, Phys. Rev. B **54**, R12653 (1996).

H. C. Xu, Y. Zhang, M. Xu, R. Peng, X.P. Shen, V.N. Strocov, M. Shi, M. Kobayashi, T. Schmitt, B. P. Xie and D. L. Feng, Phys. Rev. Lett. **112**, 087603 (2014).

N. Xu, H. Weng, B. Lv, C. Matt, J. Park, F. Bisti, V.N. Strocov, D.J. Gawryluk, E. Pomjakushina, K. Conder, N. Plumb, M. Radovic, G. Autès, O. Yazyev, Z. Fang, X. Dai, T. Qian, J. Mesot, H. Ding, M. Shi, Nature Comm. **7**, 11006 (2016).

S.-Y. Xu, N. Alidoust, I. Belopolski, Z. Yuan, G. Bian, T.-R. Chang, H. Zheng, V.N. Strocov, D.S. Sanchez, G. Chang, C. Zhang, D. Mou, Y. Wu, L. Huang, C.-C. Lee,     S.-M. Huang, B.-K. Wang, A. Bansil, H.-T. Jeng, T. Neupert, A. Kaminski, H. Lin, S. Jia & M.Z. Hasan, Nature Phys. **11**, 748 (2015-1).

S.-Y. Xu, I. Belopolski, D. S. Sanchez, C. Zhang, G. Chang, C. Guo, G. Bian, Z. Yuan, H. Lu, T.-R. Chang, P. P. Shibayev, M. L. Prokopovych, N. Alidoust, H. Zheng, C.-C. Lee, S.-M. Huang, R. Sankar, F. Chou, C.-H. Hsu, H.-T. Jeng, A. Bansil, T. Neupert, V. N. Strocov, H. Lin, S. Jia and M.Z. Hasan. Science Advances **1**, e1501092 (2015-1).

F. C. Zhang and T. M. Rice, Phys. Rev. B **37**, 3759 (1988).

Z. Zhong and P. J. Kelly, Europhys. Lett. **84**, 27001 (2008).

Z. Zhong, A. Tóth and K. Held, Phys. Rev. B **87**, 161102(R) (2013).

Z. Zhong, M. Wallerberger, J. M. Tomczak, C. Taranto, N. Parragh, A. Toschi, G. Sangiovanni and K. Held, Phys. Rev. Lett. **114**, 246401 (2015)

K. Zhou, M. Radovic, J. Schlappa, V. N. Strocov, R. Frison, J. Mesot, L. Patthey and T. Schmitt, Phys. Rev. B **83**, 201402(R) (2011)